\documentclass{article}

\usepackage[english]{babel}

\usepackage{graphicx}

\usepackage{tikz}
\usetikzlibrary{arrows.meta, positioning}

\usepackage{xcolor}

\usepackage[a4paper,top=2cm,bottom=2cm,left=3cm,right=3cm,marginparwidth=1.75cm]{geometry}

\usepackage{xcolor}
\usepackage{booktabs}
\usepackage{float}
\usepackage{url}
\usepackage{amsmath, amssymb, amsthm}
\usepackage{graphicx}
\usepackage[round, authoryear]{natbib}
\usepackage[colorlinks=true, allcolors=blue]{hyperref}
\usepackage{subcaption}
\usepackage{abstract}
\usepackage{authblk}
\usepackage{algorithm}
\usepackage{algorithmic}

\newtheorem{theorem}{Theorem}
\newtheorem{lemma}{Lemma}
\newtheorem{proposition}{Proposition}
\newtheorem{corollary}{Corollary}

\usepackage{authblk}
\usepackage{geometry}
\title{\textbf{Is Causality Necessary for Efficient Portfolios?\\ A Computational Perspective on Predictive Validity and Model Misspecification}}

\author{Alejandro Rodriguez Dominguez}
\affil{Director of Quantitative Analysis and Artificial Intelligence\\
Miralta Finance Bank S.A., Madrid, Spain\\
Email: \texttt{arodriguez@miraltabank.com}}

\date{\today}

\begin{document}

\maketitle

\begin{abstract}
Portfolio optimization is increasingly argued to require causally identified return predictors to avoid signal inversion and optimization failure. This paper re-examines this claim by studying when predictive signals yield viable efficient frontiers, even under structural misspecification. We show that causal identification is not necessary for portfolio efficiency within static mean--variance and closely related quadratic portfolio optimization frameworks. Instead, efficiency is governed by geometric sufficiency conditions on predictive signals: directional alignment, ranking preservation, and calibration. We formally decompose portfolio efficiency into these three components and show that miscalibration alone attenuates Sharpe ratios even when alignment and ranking are preserved. Robustness is characterized as smooth degradation rather than collapse, with explicit attenuation behavior and continuity of performance under increasing misspecification. The theoretical results are supported by simulations and empirical analysis. Empirical validation combines equity-based illustrations with a large global bond universe spanning multiple currencies, countries, sectors, maturities along the term structure, seniority classes, and credit ratings, together with high-dimensional stress tests, nonlinear data-generating processes, rolling-window analyses, covariance regularization, realistic portfolio constraints, and bootstrap-based statistical validation. Across these settings, optimization geometry remains well behaved whenever directional alignment is preserved. The results clarify the boundary between causality and portfolio optimization: causality may inform signal representation, but portfolio efficiency at the optimization stage is a geometric property conditional on a given representation.
\end{abstract}

\vspace{1em}

\section{Introduction}

Recent contributions in empirical finance have argued that portfolio optimization requires causal factor models, contending that structurally misspecified predictors (those that omit relevant variables or misrepresent functional relationships) necessarily generate distorted signals and collapse efficient frontiers. Within this perspective, causal identifiability is presented as a prerequisite for viability: without it, signal inversion and optimization failure are presumed unavoidable \citep{lopez2025causal}.

A recurring claim in the recent causal factor--investing literature is that improvements in predictive performance achieved through causal identification are necessary, or at least taken to be strongly diagnostic, for portfolio efficiency. This claim conflates two distinct questions. From the perspective of modern portfolio theory, efficiency is not a property of individual predictors, but of the joint geometry induced by the return signal and the risk model. Conditional on a given admissible risk--return geometry, efficiency is governed by how predictive signals align, rank, and scale within that space, rather than by whether those signals admit a causal interpretation. Causal structure may improve signal construction, but it is not intrinsic to the definition or feasibility of efficient portfolios within this framework. Throughout the paper, by quadratic portfolio optimization frameworks, we refer to static mean--variance optimization and equivalent formulations that admit a quadratic risk objective.

This paper challenges the universality of the causal-necessity claim. Through theoretical analysis, controlled simulations, and empirical evidence, we demonstrate that predictive models can remain operationally effective for portfolio construction even when structurally misspecified. In particular, we show that signals derived from such models often preserve directional alignment with the true return vector, which is sufficient to sustain convex and non-degenerate efficient frontiers under classical mean--variance optimization. Inefficiency, in practice, arises more frequently from miscalibration of signal magnitudes than from systematic sign inversion.

More generally, the results of this paper establish that causal identifiability is not a necessary condition for portfolio efficiency within the class of quadratic portfolio optimization problems studied here. Portfolio diversification is inherently a geometric problem: efficiency is determined by the relative orientation, ordering, and scaling of predictive signals within a given risk--return geometry. As shown in this work, three geometric properties—directional alignment, ranking preservation, and calibration—jointly characterize investment efficiency independently of the mechanism generating the signals.

Two clarifications are therefore essential to properly situate this contribution. First, the claim advanced here is not that causality is irrelevant, but that it is not intrinsic to the objective of static portfolio optimization. In the classical formulation of modern portfolio theory \citep{markowitz1952portfolio}, efficiency is defined geometrically through the joint distribution of returns and their covariance structure, rather than through the causal mechanisms generating those returns. Treating causal identifiability as a prerequisite for efficiency therefore conflates explanatory objectives with optimization geometry.

Second, when causality is incorporated into portfolio construction, it enters through the specification of the diversification geometry itself rather than as a requirement imposed on return prediction. Related work by \citet{RODRIGUEZDOMINGUEZ2023100447,RodriguezDominguez2025CausalPortfolioOptimization} addresses a complementary setting in which diversification is modeled through time-varying sensitivities to common causal drivers. In that framework, alignment, ranking, and calibration are evaluated relative to the induced local geometry, illustrating that even causal formulations ultimately operate through geometric conditions at the optimization stage.

The contributions of this paper are fourfold. First, we introduce and formalize a three-way distinction between directional alignment, ranking preservation, and calibration in portfolio optimization, showing that directional correctness is necessary but not sufficient for portfolio efficiency. Second, we establish geometric results demonstrating that structurally misspecified yet directionally aligned signals preserve frontier convexity and optimization viability. Third, we extend the portfolios-from-sorts framework of \citet{almgren2005sorts} by explicitly separating its ranking-based insight from other dimensions of predictive validity and by deriving a scaling law that links Sharpe ratio performance to the cosine alignment between true and surrogate signals. While \citet{almgren2005sorts} shows that correct ranking ensures membership in the efficient set, we further establish that calibration governs quantitative efficiency within that set. Fourth, we validate these results using both synthetic data and S\&P~500 equity returns, showing that predictive signals lacking causal identifiability can nonetheless support coherent and actionable efficient frontiers.

These findings contribute to the broader debate in finance and econometrics concerning the relative roles of causality and prediction. As emphasized by \citet{shmueli2010explain}, explanatory and predictive models serve distinct purposes and should be evaluated against different criteria. Recent critiques have underscored the epistemic limitations of causal reductionism in financial systems: because markets are adaptive and reflexive, unidirectional causal inference may be unstable or conceptually ill-suited \citep{polakow2024epistemic}. From this vantage, calibration and directional validity constitute more pragmatic benchmarks for assessing model utility in portfolio design than structural fidelity per se.

We do not dispute the value of causal inference, which remains indispensable for structural attribution, regime monitoring, and stress testing \citep{Wilcox2014HierarchicalCI,09641e90-d989-3dd6-88ae-92e35971022e}. Rather, we caution against overstating its necessity for optimization. What ultimately matters for portfolio construction is not causal identifiability per se, but the directional fidelity, ranking stability, and calibration of predictive signals relative to the chosen diversification geometry.

To clarify the logic of the argument and the scope of the results, Figure~\ref{fig:logic_schematic} provides a structured overview of how predictive models, geometric signal properties, and optimization interact in this paper.

Recent arguments for the necessity of causal factor modeling rely on a small number of distinct claims regarding misspecification, signal inversion, optimization failure, and fragility of efficiency under noise. For clarity, these claims and their logical refutations are summarized in Table~\ref{tab:refutation-summary} in Appendix~\ref{appendix:refutation}. The main text addresses them sequentially: Section~\ref{sec:prelim} examines bias under omitted variables, Section~\ref{sec:structural_cancellation}  refutes the generic inversion claim, Sections~\ref{sec:optundermiss} and~\ref{sec:nonlinear} characterize the conditions under which optimization geometry remains viable, and Section~\ref{sec:approx_cancellation} establishes continuity and robustness under approximate cancellation.

The remainder of the paper is structured as follows. Section~\ref{sec:background} reviews related work on causal and predictive modeling in finance. Section~\ref{sec:3} develops the theoretical framework. Section~\ref{sec:results} presents empirical validation through simulations and market data. Section~\ref{sec:discussion} examines implications for portfolio theory and modeling practice. Section~\ref{sec:conclusion} concludes.

\begin{figure}[t]
\centering
\begin{tikzpicture}[
    node distance=2.6cm,
    every node/.style={
        draw,
        rectangle,
        rounded corners,
        align=center,
        minimum width=4.2cm,
        minimum height=1.1cm,
        font=\small
    },
    arrow/.style={->, thick}
]

\node (model) {\textbf{Predictive Model}\\
{\footnotesize linear / nonlinear}\\
{\footnotesize causal or non-causal}};

\node (misspec) [below of=model] {\textbf{Structural Misspecification}\\
{\footnotesize omitted variables}\\
{\footnotesize nonlinear confounding}};

\node (signal) [right of=model, xshift=2.8cm] {\textbf{Signal Geometry}\\
{\footnotesize directional alignment}\\
{\footnotesize ranking preservation}\\
{\footnotesize calibration}};

\node (failure) [below of=signal] {\textbf{Failure Modes}\\
{\footnotesize attenuation (miscalibration)}\\
{\footnotesize inversion (special case)}};

\node (robust) [right of=signal, xshift=2.8cm] {\textbf{Geometric Robustness}\\
{\footnotesize convexity preserved}\\
{\footnotesize smooth degradation}};

\node (opt) [below of=misspec, yshift=-0.4cm] {\textbf{Portfolio Optimization}\\
{\footnotesize mean--variance geometry}\\
{\footnotesize quadratic risk}};

\node (frontier) [right of=opt, xshift=2.8cm] {\textbf{Efficient Frontier}\\
{\footnotesize feasible / convex}\\
{\footnotesize Sharpe scales with alignment}};

\node (empirical) [below of=frontier] {\textbf{Empirical Validation}\\
{\footnotesize simulations + S\&P 500}\\
{\footnotesize high-dimensional stress tests}};

\draw[arrow] (model) -- (signal);
\draw[arrow] (model) -- (misspec);
\draw[arrow] (misspec) -- (failure);
\draw[arrow] (signal) -- (failure);
\draw[arrow] (signal) -- (robust);
\draw[arrow] (robust) -- (frontier);
\draw[arrow] (opt) -- (frontier);
\draw[arrow] (failure) -- (opt);
\draw[arrow] (frontier) -- (empirical);

\end{tikzpicture}
\caption{Logical structure of the framework and the paper. Predictive models—whether causal or non-causal—produce surrogate signals whose geometric properties determine optimization behavior. Structural misspecification primarily induces attenuation through miscalibration rather than generic inversion. Portfolio efficiency is governed by alignment, ranking, and calibration within a given risk--return geometry, with robustness characterized by smooth degradation rather than frontier collapse.}
\label{fig:logic_schematic}
\end{figure}

\section{Literature Review}
\label{sec:background}

\subsection{Causal Identification and Causal Representations in Finance}
\label{sec:causal_lit}

A growing strand of recent literature in asset pricing has emphasized the role of causal factor models in portfolio construction. According to this view, models that omit causal structure (whether due to unobserved confounders, misspecified functional forms, or reliance on purely associational data) are argued to risk distorted inference and unstable optimization outcomes \citep{lopez2025causal}. This perspective is further developed in \citet{lopez2023causal,lopez2025factorMirage,lopez2025necessary}, where causal identification is presented as a safeguard against factor mirages and misleading allocation decisions.

More recently, this line of work has been extended through causal network and causal discovery approaches applied to asset pricing and factor investing. Examples include causal spillover networks across asset and factor systems \citep{KonstantinovFabozzi2026}, causal network representations for factor investing \citep{HowardLohreMudde2025}, and automated causal discovery in financial markets under nonstationarity \citep{SokolovEtAl2024,SadeghiGopalFesanghary2025,SadeghiSimonian2025}. Related contributions formalize causal inference frameworks for asset pricing \citep{HaddadEtAl2025CausalAssetPricing} and survey the rapidly expanding use of causal methods across finance and insurance \citep{KumarEtAl2025CausalBFSIReview}. Several of these works implicitly or explicitly suggest that causal identification is required to avoid systematic portfolio inefficiency, particularly in factor-based investment settings.

\subsection{Prediction, Estimation Error, and Portfolio Optimization}
\label{sec:prediction_optimization}

Yet causal-necessity arguments rest on assumptions that are often restrictive in practice. In particular, they conflate explanatory validity with the requirements of portfolio efficiency. From the standpoint of modern portfolio theory, efficiency is fundamentally a diversification concept, defined relative to the geometry induced by the asset universe and the admissible risk model, rather than by the causal provenance of return predictors.

The methodological distinction between explanation and prediction is well established. \citet{shmueli2010explain} emphasize the contrast between explanatory models, which seek causal interpretation, and predictive models, which prioritize forecasting accuracy. Similarly, \citet{breiman2001statistical} describes the divide between the "data modeling" culture oriented toward interpretability and the "algorithmic modeling" culture emphasizing predictive performance. Finance magnifies this tension. Financial markets rarely provide stable instruments, exhibit pervasive nonstationarity, and display reflexive dynamics in which causal graphs may be unstable or unobservable. In such environments, predictive validity frequently dominates causal alignment as the pragmatic standard of model utility.

Moreover, portfolio optimization is notoriously sensitive to estimation error. Seminal work by \citet{best1991sensitivity}, \citet{chopra1993errors}, and \citet{britten1999sampling} shows that even small errors in expected return estimates can induce large swings in portfolio weights. This fragility has motivated a wide range of robustification strategies, including the Black--Litterman model \citep{black1992global}, constrained optimization \citep{jagannathan2003constraints}, and shrinkage-based estimators \citep{ledoit2004honey}. Further, \citet{demiguel2009naive} demonstrate that simple heuristics, such as equal-weighted portfolios, can outperform optimized allocations when estimation error is severe.

Against this backdrop, predictive-first methodologies have gained momentum. \citet{gu2020ml} show that machine learning models, despite limited causal interpretability, deliver strong predictive performance and competitive portfolio allocations. Related evidence suggests that return-predictive signals can remain operationally effective even without explicit causal grounding \citep{haugen1996commonality, harvey2016crosssection}. In parallel, a growing literature has explored non-causal dependence measures that extend beyond correlation without invoking causal structure \citep{opdyke2021correlation, viole2019partial}. Together, these contributions demonstrate that associational and predictive modeling constitutes a viable alternative to fully specified structural approaches.

\subsection{Scope of the Critique and the Necessity Question}
\label{sec:scope_necessity}

Recent causal discovery and causal network studies primarily aim to improve interpretability, stress testing, or predictive stability rather than to analyze portfolio efficiency directly. While such methods may inform signal construction or risk attribution, they do not characterize the conditions under which efficient frontiers remain viable within a given optimization framework. In that setting, portfolio efficiency depends on how signals interact within a diversification geometry, rather than on the correctness or completeness of an underlying causal graph.

To clarify how the present contribution relates to prior work, it is useful to distinguish three broad strands in the literature. A first strand establishes that directional or ranking information may suffice for portfolio efficiency under stylized or correctly specified settings (e.g., \citet{almgren2005sorts}). A second strand documents empirically that non-causal or purely predictive signals, including machine-learning-based forecasts, can yield competitive portfolios in practice (e.g., \citet{demiguel2009naive, gu2020ml}). A third strand emphasizes the fragility of portfolio optimization under estimation error or misspecification, often motivating structurally interpretable or causal modeling approaches (e.g., \citet{best1991sensitivity, chopra1993errors}).

None of these strands, however, directly addresses the necessity question that motivates the present study: whether causal identification is required for portfolio efficiency within the static mean--variance and related optimization frameworks considered in this paper, and if not, what minimal conditions replace it. The contribution of this paper is to resolve this question by formalizing geometric sufficiency conditions based jointly on directional alignment, ranking preservation, and calibration. We show that these conditions govern portfolio efficiency independently of causal identification within this framework, persist under nonlinear data-generating processes, and degrade continuously rather than collapsing under misspecification. Accordingly, the analysis that follows is not concerned with proposing new predictors or causal discovery methods, but with clarifying the boundary between causal relevance and predictive sufficiency in portfolio optimization.

\section{Methods}
\label{sec:3}

This section develops the theoretical analysis around three claims that underlie
recent arguments for the necessity of causal modeling in portfolio optimization:
(i) that structural misspecification generically induces signal inversion;
(ii) that portfolio inefficiency arises only through such inversion; and
(iii) that efficient frontiers collapse in the absence of causal identification.
All results in this section are derived within static mean--variance and related
quadratic portfolio optimization frameworks. Sections~\ref{sec:prelim}--\ref{sec:structural_cancellation} analyze the behavior
of predictive signals under omitted variables and confounding, while
Sections~\ref{sec:optundermiss}--\ref{sec:approx_cancellation} characterize the conditions under which portfolio optimization
remains well posed within these settings.

While individual elements such as directional information, ranking, and estimation
error sensitivity have appeared separately in prior portfolio and asset pricing
literature, the contribution of this paper is to resolve an explicit necessity
question that has remained implicit: whether causal identification is required for
portfolio efficiency within the class of optimization frameworks considered here.
We provide new results that jointly characterize efficient frontier viability
under misspecification through geometric sufficiency conditions, identify
calibration as the quantitative driver of Sharpe attenuation even under correct
alignment and ranking, and show that inefficiency degrades continuously—rather than
collapsing—under nonlinear and high-dimensional misspecification. These results
are not implied by existing work that considers alignment, ranking, or estimation
error in isolation.

\subsection{Preliminary Work: Analyzing Misspecification in Linear Factor Models}
\label{sec:prelim}

This section analyzes the mathematical consequences of model misspecification on portfolio efficiency, focusing on linear factor models with omitted confounders. We examine how such structural misspecification introduces bias in estimated factor loadings and how this bias propagates to portfolio weights, even under valid optimization procedures.

\noindent\textit{Notation.} Throughout Section~\ref{sec:3}, $\mu$ denotes the true expected return vector, $\tilde{\mu}$ a surrogate or misspecified signal, $\Sigma$ the return covariance matrix, and $\omega$ portfolio weights. All inner products and norms are defined with respect to the geometry induced by $\Sigma^{-1}$ unless stated otherwise.

\subsubsection{Bias from Omitting a Confounder}

Assume asset returns follow a true structural factor model:
\[
X = B_{\text{true}} F + \varepsilon
\]
where:
\[
B_{\text{true}} =
\begin{bmatrix}
\gamma_1 & \beta_1 \\
\gamma_2 & \beta_2
\end{bmatrix}, \quad
F =
\begin{bmatrix}
Z \\
F_2
\end{bmatrix}
\]
Here, \( Z \) is an unobserved confounder that affects both the returns and the observed factor \( F_2 \). Suppose:
\[
F_2 = \delta Z + \eta, \quad \eta \sim \mathcal{N}(0, 1 - \delta^2), \quad Z \sim \mathcal{N}(0,1)
\]
with \( Z \) and \( \eta \) independent, so that\footnote{%
\citep{lopez2024case} compute this variance under an alternative normalization, reporting \(\mathrm{Var}(F_2)=1+\delta^2\), which would correspond to the specification \(\eta \sim \mathcal{N}(0,1)\). Under the definition stated in their model, \(\eta \sim \mathcal{N}(0,1-\delta^2)\), the correct result is \(\mathrm{Var}(F_2)=1\). This difference affects the scaling of the omitted-variable bias in \(\hat{\beta}_n\): under their normalization 
\(\hat{\beta}_n = (\beta_n+\gamma_n\delta)/(1+\delta^2)\), whereas under the stated normalization \(\hat{\beta}_n = \beta_n + \gamma_n \delta\). In both cases, however, omitting \(Z\) induces a $\delta$-dependent bias, so the qualitative conclusion regarding misspecification remains unchanged.}:
\[
\text{Var}(F_2) = \delta^2 \cdot \text{Var}(Z) + \text{Var}(\eta) = \delta^2 + (1 - \delta^2) = 1, \quad \text{Cov}(Z, F_2) = \delta.
\]

Now, regress \( X_n \) on \( F_2 \) alone. The estimated loading becomes:
\[
\hat{\beta}_n = \frac{\text{Cov}(X_n, F_2)}{\text{Var}(F_2)} = \text{Cov}(X_n, F_2)
\]
From the true model:
\[
X_n = \gamma_n Z + \beta_n F_2 + \varepsilon_n
\quad \Rightarrow \quad
\text{Cov}(X_n, F_2) = \gamma_n \cdot \text{Cov}(Z, F_2) + \beta_n \cdot \text{Var}(F_2) = \gamma_n \delta + \beta_n
\]

Thus:
\[
\hat{\beta}_n = \beta_n + \gamma_n \delta
\]

This shows that omitting the confounder \( Z \) introduces systematic bias into the estimated factor loading \( \hat{\beta}_n \), depending on the true confounder exposure \( \gamma_n \) and its correlation \( \delta \) with \( F_2\) \citep{lopez2024case}.

\subsubsection{Resulting Portfolio Inefficiency}
\label{sec:312}

Suppose an investor constructs a minimum-variance portfolio targeting exposure
$c^{\top} = [0,\,1]$ using the misspecified loadings
\[
B_{\text{miss}}
=
\begin{pmatrix}
\hat{\beta}_1 \\
\hat{\beta}_2
\end{pmatrix}
=
\begin{pmatrix}
\beta_1 + \gamma_1 \delta \\
\beta_2 + \gamma_2 \delta
\end{pmatrix},
\]
as in \citet{lopez2024case,lopez2025causal}. The optimizer solves\footnote{This expression implicitly assumes a two-asset setting in which the exposure constraint admits a one-dimensional null space. While not stated explicitly in \citep{lopez2024case,lopez2025causal}, this restriction is required for the above expression to be well defined.}
\[
\omega^{\ast}_{\text{miss}}
=
B_{\text{miss}}^{\top -1} c^{\top}
=
\frac{1}{\hat{\beta}_2 - \hat{\beta}_1}
\begin{pmatrix}
-1 \\
\;\,1
\end{pmatrix}.
\]

We evaluate the true exposure delivered by this portfolio under the correctly specified loading matrix
\[
B_{\text{true}}
=
\begin{pmatrix}
\gamma_1 & \beta_1 \\
\gamma_2 & \beta_2
\end{pmatrix}.
\]
A direct calculation gives
\[
\omega^{\ast\top}_{\text{miss}} B_{\text{true}}
=
\frac{1}{\hat{\beta}_2 - \hat{\beta}_1}
\left(
\gamma_2 - \gamma_1,\;
\beta_2 - \beta_1
\right).
\]

Unless $\gamma_1 = \gamma_2 = 0$, that is, unless the omitted confounder has identical exposure across assets,
the realized exposure deviates from the intended target $[0,\,1]$.
Thus, even within the work of \citet{lopez2024case,lopez2025causal},
factor misspecification alone suffices to induce portfolio inefficiency,
despite the use of a formally valid optimization procedure. This result follows purely from statistical misspecification of the factor loadings and does not rely on causal interpretation. In particular, inefficiency arises without requiring signal inversion, a distinction that will be central in the subsequent analysis.

\subsubsection{Conditions Underpinning the Necessity Claim}

The central thesis of \citet{lopez2025causal} is that causal factor modeling is a necessary condition for investment efficiency. However, this claim rests on several assumptions, some explicit, others implicit, which are listed and critically evaluated in Table~\ref{tab:causal-conditions}.

\begin{table}[H]
\centering
\renewcommand{\arraystretch}{1.2}
\small
\begin{tabular}{|p{1.5cm}|p{8.3cm}|p{2.7cm}|}
\hline
\textbf{Condition} & \textbf{Description} & \textbf{True/False} \\
\hline
\textbf{C1} & Causal factor models are necessary for portfolio efficiency. 
& \textcolor{red}{\textbf{False within the static optimization framework studied here}} — disproven by counterexample. \\
\hline
\textbf{C2} & Associational models (e.g., correlation-based) are always misspecified due to confounders or colliders. 
& \textcolor{red}{\textbf{False}} — not true if omitted variables are independent. \\
\hline
\textbf{C3} & Mean-variance optimization requires unbiased exposures $B$. 
& \textcolor{red}{\textbf{False}} — other allocation rules exist (e.g., classification-based). \\
\hline
\textbf{C4} & The data-generating process is linear and additive. 
& \textcolor{red}{\textbf{False}} — unrealistic in many real-world markets. \\
\hline
\textbf{C5} & All observed associations must be interpreted causally; correlation cannot justify predictive modeling. 
& \textcolor{red}{\textbf{False}} — contradicted by predictive theory (see \citep{shmueli2010explain}). \\
\hline
\textbf{C6} & Causal discovery tools can recover valid DAGs from financial data. 
& \textcolor{orange}{\textbf{Uncertain}} — relies on strong assumptions (e.g., no latent confounders, i.i.d.). \\
\hline
\textbf{C7} & The true causal graph is stable and estimable. 
& \textcolor{red}{\textbf{False}} — often untrue due to nonstationarity and market adaptation. \\
\hline
\end{tabular}
\caption{Conditions required for the necessity of causal modeling, as claimed in \citet{lopez2025causal}, and prior work \citet{lopez2024case,lopez2024why}.}
\label{tab:causal-conditions}
\end{table}

As shown in Table \ref{tab:causal-conditions}, the necessity claim relies on several assumptions that often do not hold in practical or general settings. If even one of these conditions fails, causal factor modeling cannot be asserted as a general requirement for investment efficiency. As our counterexamples and derivations show, Condition C1 is refuted outright, while C2–C5 are context-dependent and frequently violated in financial modeling.

The following Section introduces the structural cancellation counterexample, in which omitted variables are neutralized through bias absorption.
\subsection{Structural Cancellation as a Counterexample}
\label{sec:structural_cancellation}

This section addresses the claim that structural misspecification generically leads to signal inversion. We begin with a counterexample showing that omitted variables do not inevitably cause signal inversion.  
Instead, omitted-variable bias can produce approximate cancellation, preserving directional validity while attenuating magnitudes.  
This challenges prior claims that structural misspecification necessarily yields inverted signals and collapsed frontiers.

This counterexample establishes structural robustness to misspecification: omitted-variable bias may attenuate predictive signals without inducing systematic sign inversion, thereby preserving directional informativeness under moderate departures from the true data-generating process.

\subsubsection{Setup}

Let the observed feature be $X \sim \mathcal{N}(0,1)$ and the latent confounder be
\[
Z' = -\alpha X + \eta + \zeta, 
\quad \eta \sim \mathcal{N}(0,\sigma_\eta^2), 
\quad \zeta \sim \mathcal{N}(0,\sigma_\zeta^2),
\]
independent of $X$.  
The true signal is
\[
s_{\mathrm{true}} = \beta X + \gamma Z',
\]
and returns are generated as
\[
Y = s_{\mathrm{true}} + \varepsilon, 
\quad \varepsilon \sim \mathcal{N}(0,\sigma^2).
\]

\subsubsection{Misspecified Estimation}

Suppose we regress $Y$ on $X$ alone, omitting $Z'$.  
The OLS coefficient is
\[
\hat{\beta} = \frac{\mathbb{E}[X Y]}{\mathbb{E}[X^2]}.
\]

Substituting,
\[
\hat{\beta} = \frac{\mathbb{E}[X (\beta X + \gamma Z' + \varepsilon)]}{\mathbb{E}[X^2]}
= \beta + \gamma \cdot \frac{\mathbb{E}[X Z']}{\mathbb{E}[X^2]}.
\]

Because $Z' = -\alpha X + \eta + \zeta$, we obtain
\[
\mathbb{E}[X Z'] = -\alpha \,\mathbb{E}[X^2] = -\alpha.
\]
Thus,
\[
\hat{\beta} = \beta - \alpha \gamma.
\]

\begin{lemma}[Structural Cancellation]
\label{lem:structural_cancellation}
If $\beta$ and $\alpha \gamma$ have the same sign and comparable magnitude, the OLS coefficient
\[
\hat{\beta} = \beta - \alpha \gamma
\]
remains positive but attenuated relative to $\beta$.  
Hence, omitted-variable bias may shrink weights toward zero rather than invert them.
\end{lemma}

\begin{proof}
If $\beta>0$ and $\alpha \gamma >0$, then $\hat{\beta} = \beta - \alpha \gamma$.  
When $0 < \alpha \gamma < \beta$, we have $0 < \hat{\beta} < \beta$.  
Thus, directionality is preserved, though magnitude is reduced.  
An analogous argument applies for $\beta<0$.
\end{proof}

\begin{corollary}[Omitted-Variable Bias Does Not Necessarily Imply Inefficiency]
\label{cor:ovb_not_failure}
Under the conditions of Lemma~\ref{lem:structural_cancellation}, the misspecified estimator $\hat{\beta}$ retains the correct sign.  
Therefore, directional informativeness is preserved, and the resulting portfolio weights remain viable for convex frontier construction.  
Misspecification reduces efficiency through attenuation but does not induce systematic failure.
\end{corollary}

\noindent
\textit{Remark.} Contrary to prior claims, sign inversion requires $\alpha \gamma > \beta$, not misspecification per se.  
In practice, moderate confounding is more likely to yield attenuated but still directionally valid signals.

In econometric analysis, omitted-variable bias is typically assessed through consistency and causal interpretability, where bias implies invalid inference. In portfolio optimization, however, the relevant criterion is directional informativeness for allocation, rather than unbiased estimation for inference. The results above show that omitted-variable bias may attenuate or rescale predictive signals without destroying their alignment with realized returns, explaining why structural misspecification can invalidate causal interpretation while leaving optimization performance intact. This distinction parallels the explanatory--predictive divide emphasized by \citet{shmueli2010explain} and \citet{breiman2001statistical}, where models may be unsuitable for causal interpretation yet remain operationally effective for prediction.

Figure~\ref{fig:weights-cancellation} shows a Monte Carlo simulation of the cancellation effect.  
Despite omitting $Z'$, the estimated coefficients align in sign with the true signal but exhibit attenuation.  
This demonstrates that misspecification can preserve frontier validity when structural dependence partially cancels.

\subsection{Optimization Under Misspecification}
\label{sec:optundermiss}

This section addresses the claim that directional correctness alone is sufficient for portfolio efficiency.
We now distinguish prediction, ranking, and efficient allocation.
Throughout this section, efficiency initially refers to optimality with respect to the mean--variance objective,
and later to position along the mean--variance efficient frontier.
Our results demonstrate that directional agreement between signals is necessary but not sufficient for portfolio efficiency:
signals may align in sign yet still yield inefficient allocations due to miscalibration.
This refines prior claims that inefficiency arises only through sign inversion.

\begin{theorem}[Directional Agreement Does Not Imply Efficiency]
\label{thm:alignment_efficiency}
There exist predictive signals $\omega_{\mathrm{pred}}$ such that 
\[
\operatorname{sign}(\omega_{\mathrm{pred}}(x))=\operatorname{sign}(\omega_{\mathrm{true}}(x)) 
\quad \forall x\in\mathcal{X},
\]
yet
\[
R(\omega_{\mathrm{pred}}) \neq R(\omega_{\mathrm{true}}),
\quad
J(\omega_{\mathrm{pred}}) < J(\omega_{\mathrm{true}}),
\]
where $R(\omega)=\mathbb{E}[\omega(x)r(x)]$ and 
\[
J(\omega)=\mathbb{E}[\omega(x)r(x)]-\lambda\,\mathrm{Var}(\omega(x)r(x)),\qquad \lambda>0.
\]

\end{theorem}

\begin{proof}
Let $\omega_{\mathrm{pred}}(x)=\epsilon\,\omega_{\mathrm{true}}(x)$ with $\epsilon \in (0,1)$.  
Then
\[
R(\omega_{\mathrm{pred}}) = \epsilon R(\omega_{\mathrm{true}})\neq R(\omega_{\mathrm{true}}).
\]
Moreover, letting $Y=\omega_{\mathrm{true}}(x)r(x)$,
\[
J(\omega_{\mathrm{pred}})
=\mathbb{E}[\epsilon Y]-\lambda\,\mathrm{Var}(\epsilon Y)
=\epsilon\,\mathbb{E}[Y]-\lambda\,\epsilon^2\,\mathrm{Var}(Y)
<\mathbb{E}[Y]-\lambda\,\mathrm{Var}(Y)
=J(\omega_{\mathrm{true}}),
\]
since $\epsilon\in(0,1)$ and $\lambda>0$.
The case $\epsilon<0$ recovers full inversion, a special case previously emphasized.
\end{proof}

\noindent
\textit{Remark.}
Directional correctness ensures that signs match, but without proper scaling,
mean--variance efficiency contracts.

\begin{lemma}[Ranking Does Not Guarantee Efficient Sizing]
\label{lem:ranking_not_sizing}
Let $f_{\mathrm{true}},f_{\mathrm{pred}}:\mathcal{X}\to\mathbb{R}$ define weights 
$\omega_{\mathrm{true}}(x)=f_{\mathrm{true}}(x)$, 
$\omega_{\mathrm{pred}}(x)=f_{\mathrm{pred}}(x)$.  
Suppose rankings agree:
\[
f_{\mathrm{true}}(x_i)>f_{\mathrm{true}}(x_j)
\iff
f_{\mathrm{pred}}(x_i)>f_{\mathrm{pred}}(x_j).
\]
If $f_{\mathrm{pred}}$ is not a positive multiple of $f_{\mathrm{true}}$, then
\[
\mathbb{E}\big[(\omega_{\mathrm{pred}}(x)-\omega_{\mathrm{true}}(x))^2\big]>0,
\]
so efficiency is lost through miscalibration.
\end{lemma}

\begin{proof}
If $f_{\mathrm{pred}}=a f_{\mathrm{true}}$ for some $a>0$, weights differ only by scale and efficiency is preserved.
Otherwise, the deviation $\delta(x)=f_{\mathrm{pred}}(x)-f_{\mathrm{true}}(x)$ is nonzero with positive probability,
hence $\mathbb{E}[\delta(x)^2]>0$.
This implies inefficient sizing under mean--variance optimization.
\end{proof}

\begin{corollary}[Calibration Is Necessary for Quantitative Efficiency Within the Efficient Set]
\label{cor:calibration_necessary}
Even if signals preserve direction and ranking, portfolios formed from miscalibrated magnitudes
are strictly less efficient than those based on correctly scaled signals.
\end{corollary}

\begin{proof}
Follows directly from Lemma~\ref{lem:ranking_not_sizing}:
unless $f_{\mathrm{pred}}=a f_{\mathrm{true}}$ with $a>0$,
deviations in magnitude reduce realized efficiency.
\end{proof}

\noindent
\textit{Remark.}
Ranking ensures membership in the efficient set,
but calibration determines the investor’s position along the frontier.

\begin{proposition}[Sharpe Ratio with Surrogate Signals]
\label{prop:sharpe_cosine}
Let $\mu$ be the true mean return vector,
$\tilde{\mu}$ a surrogate preserving its order,
and $V\succ0$ the covariance matrix.
Portfolio weights are formed as $V^{-1}\tilde{\mu}$,
while performance is evaluated under the true mean $\mu$.
Then
\[
\mathrm{Sharpe}(V^{-1}\tilde{\mu})
=\frac{\mu^\top V^{-1}\tilde{\mu}}{\sqrt{\tilde{\mu}^\top V^{-1}\tilde{\mu}}}
=\|\mu\|_{V^{-1}}\cdot \rho_{V^{-1}},
\]
where
\[
\rho_{V^{-1}}
=\frac{\mu^\top V^{-1}\tilde{\mu}}
{\|\mu\|_{V^{-1}}\|\tilde{\mu}\|_{V^{-1}}}
\in[-1,1].
\]
\end{proposition}

\begin{proof}
We work under the standard admissibility conditions of mean--variance optimization: $V \succ 0$ and surrogate expected returns $\tilde{\mu}$ that preserve the componentwise ordering of the true mean return vector $\mu$. Under these conditions, \citet{almgren2005sorts} show that $V^{-1}\tilde{\mu}$ lies in the efficient cone.
Computing return and variance under the $V^{-1}$ geometry yields the stated scaling law.
\end{proof}

Proposition~\ref{prop:sharpe_cosine} differs from standard mean--variance results
in that the expected return vector is not assumed to be correctly specified.
Classical treatments implicitly treat $\mu$ as a primitive object,
whereas the present result characterizes portfolio efficiency
when $\tilde{\mu}$ is an informational surrogate whose alignment with $\mu$
may vary continuously.
This distinction is central to the causal-necessity debate:
efficiency does not hinge on structural correctness,
but on the geometric relation between surrogate and true signals.
This attenuation effect is consistent with the well-documented sensitivity
of mean--variance portfolios to expected return estimation error
\citep{best1991sensitivity, chopra1993errors},
but differs in that it characterizes efficiency loss geometrically rather than statistically.

The linear dependence of the Sharpe ratio on cosine alignment implies performance robustness:
efficiency degrades smoothly with declining alignment rather than collapsing abruptly under misspecification.

\begin{lemma}[Ranking--Calibration Decomposition]
\label{lem:decomposition}
Ranking is necessary for efficient-set membership,
while calibration governs quantitative efficiency within the set.
Formally: if ranking fails, efficiency is unattainable;
if ranking holds, Sharpe scales linearly with $\rho_{V^{-1}}$.
\end{lemma}

\begin{proof}
Necessity: if ordering is violated, $\tilde{\mu}$ may exit the efficient cone,
making optimization invalid.  
Sufficiency: if ranking holds,
Proposition~\ref{prop:sharpe_cosine} shows
$\mathrm{Sharpe}(V^{-1}\tilde{\mu})
=\|\mu\|_{V^{-1}}\rho_{V^{-1}}$,
so calibration (alignment $\rho_{V^{-1}}$) determines efficiency.
\end{proof}

\noindent
\textit{Remark.}
Directional correctness is necessary but insufficient;
ranking secures frontier validity,
and calibration dictates realized Sharpe.
This decomposition explains why misspecified models underperform
via attenuation rather than systematic inversion.

\subsection{Testing Optimization Geometry Under Misspecification}
\label{sec:nonlinear}

This section addresses the claim that structural misspecification necessarily collapses portfolio
optimization geometry, yielding degenerate or ill-defined efficient frontiers. Using nonlinear confounded data-generating processes, we examine whether misspecification
generically invalidates optimization.
We show that surrogate signals, even when estimated from misspecified models, can sustain
smooth and non-degenerate efficient frontiers provided they remain directionally aligned with
the true return vector.

The relevance of nonlinear data-generating processes in this context lies not in functional form
per se, but in their effect on directional alignment. Nonlinear transformations and interactions
may distort magnitudes or induce heteroskedasticity, yet they do not generically imply sign
reversal. As long as predictive models preserve monotone relationships with expected returns,
the geometric conditions underpinning optimization (directional alignment and ranking
preservation) remain intact. This perspective is consistent with empirical evidence that
predictive models can outperform structurally motivated ones in high-dimensional and nonlinear
settings \citep{demiguel2009naive, gu2020ml}.

\subsubsection{Nonlinear Confounded Signals}

Consider observable features $X_1, X_2 \sim \mathcal{N}(0,1)$ and an unobserved confounder
\[
Z = \alpha_1 X_1 + \alpha_2 X_2 + \eta, \qquad \eta \sim \mathcal{N}(0,1).
\]
Let the true signal follow the nonlinear model
\[
\mathbb{P}(Y=1 \mid X, Z)
=
\sigma\!\left(\tanh(X) + 0.5 \sin(Z)\right),
\]
with portfolio weights
\[
\omega_{\mathrm{true}} = 2\,\mathbb{P}(Y=1 \mid X,Z) - 1.
\]
A misspecified logistic regression omitting $Z$ produces surrogate signals
\[
\omega_{\mathrm{pred}} = 2\,\sigma(\hat{\beta}^\top X) - 1.
\]
This setup extends the structural cancellation mechanism: omitted-variable bias and nonlinearity
need not eliminate directional informativeness.

\subsubsection{Robustness of Optimization Geometry}

We formalize the claim that directional alignment, not structural fidelity, secures valid frontiers.

\begin{lemma}[Directional Signals Yield Valid Frontiers]
\label{lem:directional-frontier}
Let $\mu$ denote true expected returns, $\tilde{\mu}$ surrogate expected returns, and
$\Sigma \succ 0$ the covariance matrix. Consider the Markowitz problem
\[
\max_{\omega \in \mathbb{R}^n}
\;\tilde{\mu}^\top \omega - \lambda\,\omega^\top \Sigma \omega
\quad\text{s.t.}\quad \mathbf{1}^\top \omega = 1, \ \lambda>0.
\]
If $\tilde{\mu}^\top \mu > 0$, then the efficient frontier traced by
\[
\mathcal{F}(\lambda)
=
\big(\sigma(\omega^*(\lambda)), \ \mu^\top \omega^*(\lambda)\big)
\]
is smooth, convex, and non-degenerate in the $(\sigma, \mathbb{E}[r])$ plane.
\end{lemma}

\begin{proof}
The Lagrangian first-order condition gives
\[
\omega^*(\lambda)
=
\frac{1}{2\lambda}\Sigma^{-1}(\tilde{\mu}-\nu\mathbf{1}),
\]
with $\nu$ uniquely determined by the budget constraint. Since $\Sigma \succ 0$,
$\omega^*(\lambda)$ is unique and varies smoothly with $\lambda$.
The realized return is
\[
R(\lambda) = \mu^\top \omega^*(\lambda) \propto \mu^\top \Sigma^{-1} \tilde{\mu}.
\]
If $\tilde{\mu}^\top \mu>0$, then $R(\lambda)>0$ for all $\lambda>0$.
Portfolio variance $\sigma^2(\lambda)=\omega^{*\top}\Sigma\omega^*$ is strictly convex
in $\lambda$, so the mapping $\lambda\mapsto(\sigma(\lambda),R(\lambda))$ is smooth
and convex. Non-degeneracy follows from $R(\lambda)>0$.
\end{proof}

\noindent
\textit{Remark.}
The expression $\Sigma^{-1}\tilde{\mu}$ should be interpreted as an efficient
\emph{direction}. The budget constraint $\mathbf{1}^\top \omega = 1$ is enforced
by the affine normalization encoded by the Lagrange multiplier $\nu$, which ensures
$\mathbf{1}^\top \omega^*(\lambda)=1$ and does not affect frontier geometry.

\begin{corollary}[Frontier Preservation Under Misspecification]
\label{cor:frontier_preservation}
Suppose $\tilde{\mu}$ is estimated from a misspecified logistic regression omitting
confounders. If $\tilde{\mu}^\top \mu>0$, then the Markowitz frontier remains smooth,
convex, and non-degenerate.
\end{corollary}

\begin{proof}
Immediate from the previous lemma, with $\tilde{\mu}$ interpreted as a surrogate mean
vector. Directional alignment ensures positive realized returns, convexity, and smooth
variation of $\omega^*(\lambda)$.
\end{proof}

\noindent
\textit{Remark.}
Structural misspecification does not collapse optimization geometry. Directional
informativeness suffices to preserve frontier viability. In this sense, robustness refers
to preservation of optimization geometry under misspecified inputs, not to invariance
of optimal weights or realized returns.

\subsubsection{Sharpe Ratio Sensitivity to Alignment}

We next quantify how efficiency contracts as alignment weakens.

\begin{proposition}[Sharpe Ratio Scales with Alignment]
\label{prop:sharpe_alignment_scaling}
Let $\mu \in \mathbb{R}^n$ denote true expected returns and let $\nu\perp\mu$ with
$\|\nu\|=\|\mu\|$. Define
\[
\tilde{\mu}(\theta)=\cos\theta\,\mu+\sin\theta\,\nu,\qquad \theta\in[0,\pi].
\]
For $\omega^*(\theta)=\frac{1}{2\lambda}\Sigma^{-1}\tilde{\mu}(\theta)$, the realized
Sharpe ratio satisfies
\[
\mathrm{Sharpe}(\omega^*(\theta))
=
\cos\theta\cdot \mathrm{Sharpe}(\omega^*(0)).
\]
\end{proposition}

\begin{proof}
We have
\[
\mu^\top \omega^*(\theta)
=
\frac{1}{2\lambda}\mu^\top \Sigma^{-1}\tilde{\mu}(\theta)
=
\cos\theta\cdot \mu^\top \omega^*(0),
\]
since $\mu^\top \Sigma^{-1}\nu=0$. Portfolio volatility
\[
\sigma(\omega^*(\theta))
=
\frac{1}{2\lambda}\sqrt{\tilde{\mu}(\theta)^\top \Sigma^{-1}\tilde{\mu}(\theta)}
\]
is invariant in $\theta$ by construction. Dividing return by volatility yields the
stated scaling.
\end{proof}

\begin{corollary}[Frontier Collapse Under Misalignment]
\label{cor:frontier_collapse}
If $\cos\theta\le 0$, then $\mathrm{Sharpe}(\omega^*(\theta))\le 0$. Thus:
\begin{itemize}
\item $\cos\theta=0$: the portfolio is orthogonal to $\mu$ and achieves zero return;
\item $\cos\theta<0$: the portfolio is anti-aligned and delivers negative performance.
\end{itemize}
Positive directional alignment is therefore necessary for viable optimization.
\end{corollary}

\noindent
\textit{Remark.}
Efficiency degrades linearly with cosine alignment $\rho=\cos\theta$. Collapse occurs
only at $\rho\le 0$, confirming that miscalibration, not inversion, is the dominant
inefficiency channel.

\subsection{Approximate Cancellation and Attenuation}
\label{sec:approx_cancellation}

This section addresses the claim that portfolio efficiency is fragile to misspecification, in the sense that small departures from the true data-generating process necessarily induce abrupt performance collapse. We study omitted‐variable bias when a confounder nearly cancels the observed signal but is corrupted by independent noise. Let \(X\sim \mathcal{N}(0,1)\), and let
\[
Z' \;=\; -\alpha\,X + \eta + \zeta,
\qquad
\eta \sim \mathcal{N}(0,\sigma_\eta^2),\ \ \zeta \sim \mathcal{N}(0,\sigma_\zeta^2),
\]
with \(\eta,\zeta\) independent of \(X\) and of each other. Define the \emph{normalized} confounder channel
\[
\widetilde{Z} \;:=\; \frac{Z'}{\sqrt{1+\sigma_\eta^2+\sigma_\zeta^2}},
\]
so that \(\operatorname{Var}(\widetilde{Z}) = \operatorname{Var}(Z')/(1+\sigma_\eta^2+\sigma_\zeta^2)=1\). Consider the structural outcome
\[
Y \;=\; \beta\,X + \gamma\,\widetilde{Z} + \varepsilon,\qquad \mathbb{E}[\varepsilon\mid X,\widetilde{Z}] = 0,\ \ \operatorname{Var}(\varepsilon)=\sigma_\varepsilon^2,
\]
and suppose a practitioner fits a misspecified regression of \(Y\) on \(X\) only. Denote by \(\widehat\beta_{\text{OLS}}(\sigma_\zeta)\) the population OLS coefficient of \(Y\) on \(X\) obtained under this misspecification when the confounding‐noise level is \(\sigma_\zeta\).

The normalization of \(Z'\) into \(\widetilde{Z}\) is natural in practice: the structural channel (the loading \(\gamma\)) acts on a confounder with unit variance. As the extraneous noise \(\zeta\) increases, the correlation between \(X\) and \(\widetilde{Z}\) \emph{attenuates} smoothly, producing a continuous bias‐decay law that we derive exactly below.

\vspace{0.5em}
\begin{proposition}[Exact attenuation law under approximate cancellation]
\label{prop:attenuation}
With the setup above (and \(\operatorname{Var}(X)=1\)), the misspecified OLS slope is
\begin{equation}
\widehat\beta_{\emph{OLS}}(\sigma_\zeta)
\;=\;
\beta \;-\; \frac{\alpha\,\gamma}{\sqrt{1+\sigma_\eta^2+\sigma_\zeta^2}}.    
\label{eq:attenuation}
\end{equation}

In particular, as the confounding noise increases,
\[
\lim_{\sigma_\zeta\downarrow 0}\widehat\beta_{\emph{OLS}}(\sigma_\zeta) \;=\; \beta \;-\; \frac{\alpha\gamma}{\sqrt{1+\sigma_\eta^2}},
\qquad
\lim_{\sigma_\zeta\uparrow \infty}\widehat\beta_{\emph{OLS}}(\sigma_\zeta) \;=\; \beta,
\]
and \(\widehat\beta_{\emph{OLS}}(\sigma_\zeta)\) is strictly increasing in \(\sigma_\zeta\) whenever \(\alpha\gamma>0\).
\end{proposition}

\begin{proof}
Since \(\operatorname{Var}(X)=1\), population OLS gives
\(
\widehat\beta_{\text{OLS}} = \frac{\operatorname{Cov}(X,Y)}{\operatorname{Var}(X)} = \operatorname{Cov}(X,Y).
\)
Using \(Y=\beta X + \gamma \widetilde{Z} + \varepsilon\) and \(\mathbb{E}[\varepsilon\mid X,\widetilde{Z}]=0\),
\[
\operatorname{Cov}(X,Y)
= \beta\,\operatorname{Var}(X) + \gamma\,\operatorname{Cov}(X,\widetilde{Z})
= \beta + \gamma\,\operatorname{Cov}\!\Big(X,\frac{Z'}{\sqrt{1+\sigma_\eta^2+\sigma_\zeta^2}}\Big).
\]
Because \(Z'=-\alpha X + \eta + \zeta\) with \(\eta,\zeta\) independent of \(X\),
\(
\operatorname{Cov}(X,Z') = -\alpha\,\operatorname{Var}(X) = -\alpha.
\)
Hence
\[
\operatorname{Cov}(X,\widetilde{Z})
= \frac{\operatorname{Cov}(X,Z')}{\sqrt{1+\sigma_\eta^2+\sigma_\zeta^2}}
= -\frac{\alpha}{\sqrt{1+\sigma_\eta^2+\sigma_\zeta^2}}.
\]
Substituting yields the formula in the statement. The limits as \(\sigma_\zeta\) tends to 0 or \(\infty\) are immediate, and monotonicity for \(\alpha\gamma>0\) follows from differentiating w.r.t.\ \(\sigma_\zeta\):
\[
\frac{\partial \widehat\beta_{\text{OLS}}}{\partial \sigma_\zeta}
= \alpha\gamma \,\frac{\sigma_\zeta}{\big(1+\sigma_\eta^2+\sigma_\zeta^2\big)^{3/2}}
\ \ge\ 0,
\]
with strict positivity for \(\sigma_\zeta>0\).
\end{proof}

\vspace{0.25em}
\begin{corollary}[Continuity, smoothness, and bounds]
\label{cor:smoothness}
The map \(\sigma_\zeta \mapsto \widehat\beta_{\emph{OLS}}(\sigma_\zeta)\) is real‐analytic on \([0,\infty)\). Moreover,
\[
\beta - \frac{|\alpha\gamma|}{\sqrt{1+\sigma_\eta^2}}
\ \le \
\widehat\beta_{\emph{OLS}}(\sigma_\zeta)
\ \le \
\beta,
\qquad \forall\,\sigma_\zeta\ge 0,
\]
with the lower (upper) bound attained at \(\sigma_\zeta=0\) (\(\sigma_\zeta\to\infty\)) when \(\alpha\gamma>0\).
\end{corollary}

\begin{proof}
Real‐analyticity follows because \(\widehat\beta_{\text{OLS}}\) is a composition of analytic functions of \(\sigma_\zeta\). The bounds are the minimum/maximum of the strictly increasing function in Proposition~\ref{prop:attenuation} (for \(\alpha\gamma>0\)); the reversed inequalities hold when \(\alpha\gamma<0\).
\end{proof}

\vspace{0.25em}
\begin{corollary}[Non‐inversion under approximate cancellation]
\label{cor:noinversion}
If \(\beta>0\) and \(\alpha\gamma \le \beta\sqrt{1+\sigma_\eta^2}\), then
\[
\widehat\beta_{\emph{OLS}}(\sigma_\zeta) \ \ge \ \beta - \frac{\alpha\gamma}{\sqrt{1+\sigma_\eta^2}} \ \ge \ 0
\qquad \forall\,\sigma_\zeta\ge 0.
\]
Hence sign inversion cannot occur for any \(\sigma_\zeta\ge 0\). Analogous conditions hold for \(\beta<0\).
\end{corollary}

\begin{proof}
Combine the lower bound in Corollary~\ref{cor:smoothness} with the assumption \(\alpha\gamma \le \beta\sqrt{1+\sigma_\eta^2}\) when \(\beta>0\).
\end{proof}

\paragraph{Implications for signal alignment and Sharpe.}
Let \(w(\nu) \propto V^{-1}\nu\) denote the mean–variance direction for a signal \(\nu\), and evaluate performance under the \emph{true} mean \(\mu\).
For any surrogate \(\tilde\mu\), 
\[
\mathrm{Sharpe}\big(V^{-1}\tilde\mu\big)
= \frac{\mu^\top V^{-1}\tilde\mu}{\sqrt{\tilde\mu^\top V^{-1}\tilde\mu}}
= \|\mu\|_{V^{-1}}\,\rho_{V^{-1}}(\mu,\tilde\mu),
\]
with \(\rho_{V^{-1}}\in[-1,1]\). If \(\tilde\mu(\sigma_\zeta) = a(\sigma_\zeta)\,\mu + b(\sigma_\zeta)\,\nu\) where \(\nu\perp_{V^{-1}}\mu\) and \(a,b\) are continuous in \(\sigma_\zeta\) (as induced by the scalar attenuation in Proposition~\ref{prop:attenuation}), then \(\rho_{V^{-1}}(\mu,\tilde\mu(\sigma_\zeta)) = \frac{a(\sigma_\zeta)}{\sqrt{a(\sigma_\zeta)^2 + b(\sigma_\zeta)^2}}\) is continuous and strictly increasing in any regime where \(a(\sigma_\zeta)\) increases while \(b(\sigma_\zeta)\) is weakly decreasing. Consequently, \(\mathrm{Sharpe}(V^{-1}\tilde\mu(\sigma_\zeta))\) varies \emph{continuously} with \(\sigma_\zeta\) and contracts smoothly toward zero alignment as \(a(\sigma_\zeta)\downarrow 0\) (or expands as \(a(\sigma_\zeta)\uparrow 1\)).

\vspace{0.25em}
\begin{proposition}[Continuity and Lipschitz control of Sharpe under attenuation]
\label{prop:sharpe_continuity}
Suppose \(\tilde\mu(\sigma_\zeta)\) is constructed by attenuating the confounding component as in Proposition~\ref{prop:attenuation} (componentwise, with independent channels) and \(\|\tilde\mu(\sigma_\zeta)\|_{V^{-1}}\) is uniformly bounded away from zero. Then
\[
\Big|\mathrm{Sharpe}\big(V^{-1}\tilde\mu(\sigma_\zeta)\big) - \mathrm{Sharpe}\big(V^{-1}\tilde\mu(\sigma_\zeta')\big)\Big|
\ \le\ L\,|\sigma_\zeta-\sigma_\zeta'|
\]
for some \(L<\infty\) depending on \((\mu,V^{-1})\) and the attenuation profiles; in particular, the Sharpe function is continuous (indeed locally Lipschitz) in \(\sigma_\zeta\).
\end{proposition}

\begin{proof}
Write \(S(\sigma) = \frac{\mu^\top V^{-1}\tilde\mu(\sigma)}{\sqrt{\tilde\mu(\sigma)^\top V^{-1}\tilde\mu(\sigma)}}\).
By the quotient rule and Cauchy–Schwarz in the \(V^{-1}\) inner product,
\[
|S'(\sigma)|
\ \le\
\frac{\|\mu\|_{V^{-1}}\ \|\tilde\mu'(\sigma)\|_{V^{-1}}}{\|\tilde\mu(\sigma)\|_{V^{-1}}}
\;+\;
\frac{\|\mu\|_{V^{-1}}\ \|\tilde\mu(\sigma)\|_{V^{-1}}\ \|\tilde\mu'(\sigma)\|_{V^{-1}}}{\|\tilde\mu(\sigma)\|_{V^{-1}}^3}
\ \le\ C\,\|\tilde\mu'(\sigma)\|_{V^{-1}},
\]
for a finite constant \(C\) whenever \(\|\tilde\mu(\sigma)\|_{V^{-1}}\) is bounded away from zero. The attenuation map \(\sigma\mapsto \tilde\mu(\sigma)\) is smooth in each component (by Proposition~\ref{prop:attenuation}), hence \(\|\tilde\mu'(\sigma)\|_{V^{-1}}\) is bounded on compact intervals, giving the Lipschitz bound by the mean value theorem.
\end{proof}

The attenuation law \(\widehat\beta_{\text{OLS}}(\sigma_\zeta) = \beta - \alpha\gamma/\sqrt{1+\sigma_\eta^2+\sigma_\zeta^2}\) shows that omitted-variable bias decays \emph{smoothly} as idiosyncratic noise in the confounder increases. Under mild conditions, this prevents sign inversion (Corollary~\ref{cor:noinversion}) and ensures that optimization performance (\(\mathrm{Sharpe}(V^{-1}\tilde\mu)\)) varies continuously (Proposition~\ref{prop:sharpe_continuity}). Thus, signal viability is not a knife--edge property: degradation under misspecification is continuous and quantitatively controlled.

We emphasize that robustness to misspecification is quantified explicitly rather than asserted qualitatively. Proposition~\ref{prop:attenuation} and Corollaries~\ref{cor:smoothness}--\ref{cor:noinversion} establish a continuous and bounded attenuation of predictive signals as misspecification increases, while Proposition~\ref{prop:sharpe_continuity} shows that the resulting portfolio performance degrades smoothly through directional misalignment. Together, these results rule out threshold effects or knife-edge behavior. As a consequence, no minimum cosine similarity is required for viability; instead, efficiency degrades smoothly under increasing misspecification.

\section{Numerical Results}
\label{sec:results}

We now provide empirical validation of the theoretical framework developed in Section~\ref{sec:3}, focusing on the geometry of portfolio optimization under structural misspecification. The experiments are organized in five parts. First, simulations illustrate structural cancellation, showing that omitted-variable bias may attenuate but not invert predictive signals (Corollary~\ref{cor:ovb_not_failure}). Second, calibration--ranking experiments validate that ranking ensures efficient-set membership, while calibration governs quantitative efficiency (Lemma~\ref{lem:decomposition}, Proposition~\ref{prop:sharpe_cosine}). Third, nonlinear counterexamples test the robustness of optimization geometry, demonstrating that misspecification does not collapse the efficient frontier (Lemma~\ref{lem:directional-frontier}, Corollary~\ref{cor:frontier_preservation}). Fourth, alignment experiments confirm that Sharpe ratios contract linearly with cosine alignment, as predicted by Proposition~\ref{prop:sharpe_alignment_scaling} and Corollary~\ref{cor:frontier_collapse}). Finally, empirical validation using S\&P~500 equity data shows that predictive signals derived from misspecified models can sustain convex and coherent frontiers in practice.

Appendix~\ref{appendix:simulation} contains detailed descriptions of the simulation protocol, data preprocessing, normalization choices, and additional robustness checks. These materials support the empirical results in this section while keeping the main exposition focused on interpretation. Additional robustness analyses, including high-dimensional stress tests, rolling-window evaluations, sensitivity checks under alternative dependence and noise regimes, and constraint-aware optimization, are reported in Appendix~\ref{app:highD}. Complementing the equity-based empirical analysis presented in this section,
Appendix~\ref{app:highD} further reports a large-scale empirical validation using a global bond universe comprising 1{,}350 instruments across multiple currencies, countries, sectors, maturities along the term structure, seniority classes, and credit ratings, observed over 1{,}061 trading days (from Bloomberg). These experiments extend the theoretical predictions to heterogeneous risk profiles, dependence structures, and realistic portfolio constraints beyond equity markets.

Together, these results disentangle the distinct roles of directionality, ranking, and calibration in determining portfolio outcomes. All simulation code is available in the supplementary material. Algorithm~\ref{alg:simulation_procedure} summarizes the simulation design used throughout this section, highlighting how alignment, ranking, and calibration of predictive signals affect mean--variance portfolio optimization under structural misspecification. The simulations in this section examine empirical robustness by assessing whether directional alignment, calibration effects, and optimization geometry persist under increasing misspecification and noise.

\begin{algorithm}[H]
\caption{Simulation procedure for assessing predictive signal viability}
\label{alg:simulation_procedure}
\begin{algorithmic}[1]
\STATE Generate true expected returns $\mu \in \mathbb{R}^N$ and covariance matrix $\Sigma \in \mathbb{R}^{N \times N}$
\STATE Construct a surrogate predictive signal $\hat{\mu}$ under controlled misspecification (e.g., omitted variables or nonlinear transformations)
\STATE Compute directional alignment $\cos(\hat{\mu}, \mu)$ and assess ranking preservation between $\hat{\mu}$ and $\mu$
\STATE Solve the mean--variance optimization problem using $\hat{\mu}$ and $\Sigma$ to obtain portfolio weights $w(\hat{\mu})$
\STATE Evaluate the resulting efficient frontier and realized performance metrics (return, volatility, Sharpe ratio)
\STATE Repeat across varying degrees of misspecification and dependence to assess robustness
\end{algorithmic}
\end{algorithm}

\subsection{Structural Cancellation Under Misspecification}

\begin{figure}[t!]
\centering
\includegraphics[width=0.6\textwidth]{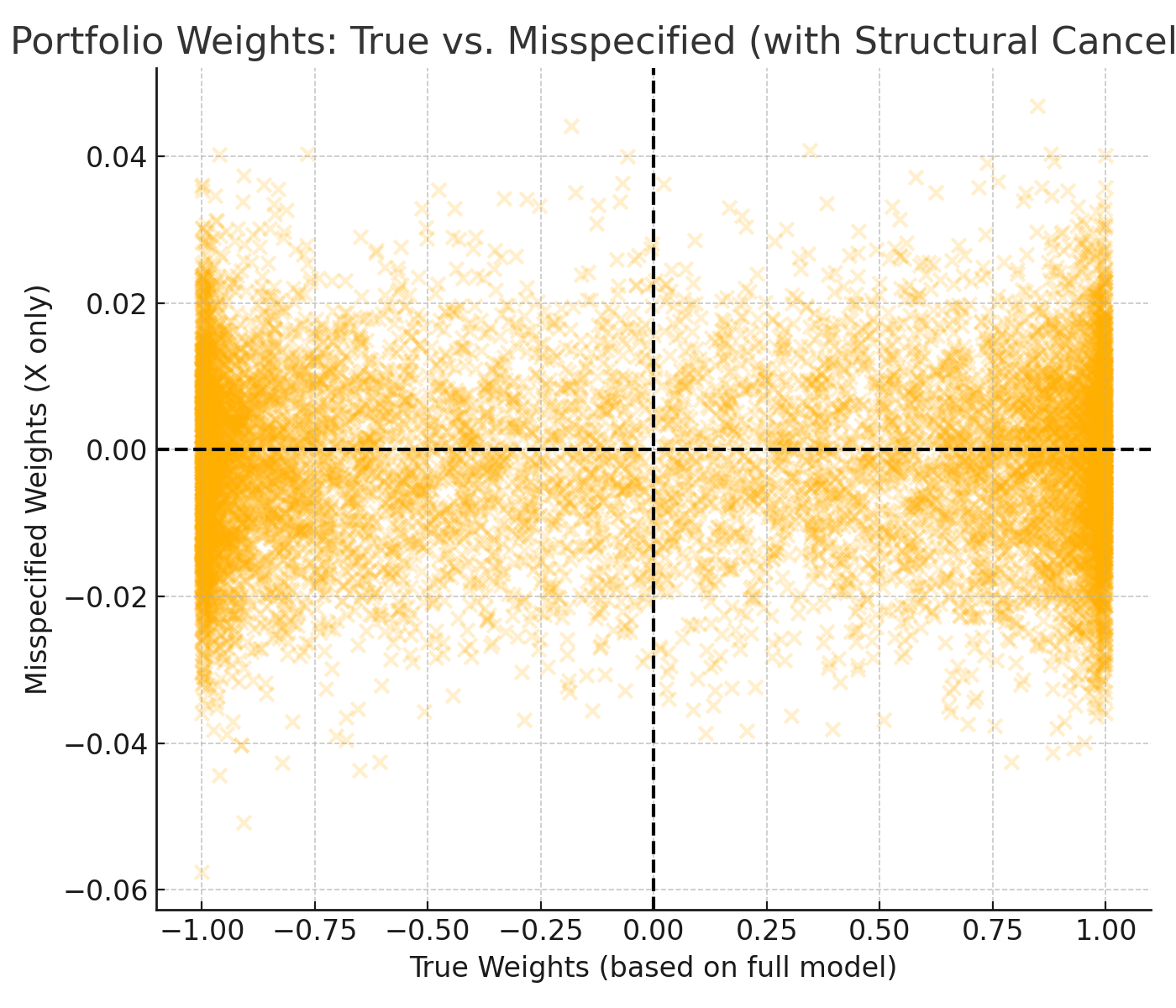}
\caption{
\textbf{Portfolio Weights under Structural Cancellation.}
Scatter of optimal portfolio weights obtained from the true expected return vector (x-axis) against weights obtained from a structurally misspecified signal (y-axis). Each point corresponds to an asset. The preservation of sign across weights indicates directional alignment, while attenuation of magnitudes reflects miscalibration. This validates that misspecification need not invalidate diversification, but primarily reduces achievable risk-adjusted returns.}

\label{fig:weights-cancellation}
\end{figure}

We simulate \( n = 1000 \) observations with features \( X \sim \mathcal{N}(0,1) \) and define a confounder \( Z' = -\alpha X + \eta \), where \( \alpha = 0.6 \) and \( \eta \sim \mathcal{N}(0, 1-\alpha^2) \). Outcomes depend on both \( X \) and \( Z' \), but the predictive model uses only \( X \). Portfolio weights are computed as
\[
\omega^{\mathrm{true}} = 2 \cdot \mathbb{P}(Y=1 \mid X,Z') - 1, 
\quad 
\omega^{\mathrm{pred}} = 2 \cdot \mathbb{P}(Y=1 \mid X) - 1.
\]
Figure~\ref{fig:weights-cancellation} shows strong directional alignment without sign inversions, confirming Lemma~\ref{lem:structural_cancellation}. Magnitudes are attenuated, but the frontier remains viable, illustrating Corollary~\ref{cor:ovb_not_failure}.

\subsection{Calibration and Ranking within the Efficient Set}
\label{sec:exp_calibration}

\begin{figure}[t!]
  \centering
  \begin{subfigure}[t]{0.49\textwidth}
    \centering
    \includegraphics[width=\linewidth]{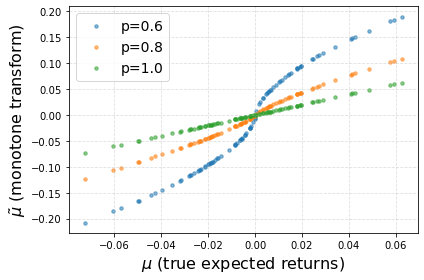}
    \caption{\textbf{Ranking preservation under monotone transformations.} Asset ranks induced by predictive signals after monotone transformations with parameters $p=0.6, 0.8, 1.0$. The x-axis reports asset rank under the true signal and the y-axis under the transformed signal. Exact rank preservation across transformations implies efficient-set membership, illustrating that feasibility depends on ordering rather than magnitude accuracy.
    }
    \label{fig:ranking_preserved}
  \end{subfigure}\hfill
  \begin{subfigure}[t]{0.49\textwidth}
    \centering
    \includegraphics[width=\linewidth]{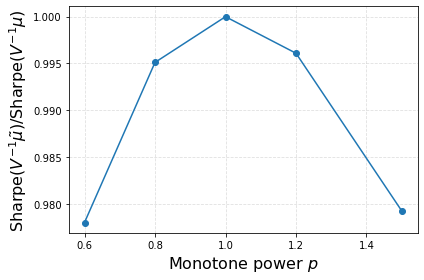}
    \caption{\textbf{Effect of calibration on risk-adjusted performance.} Population Sharpe ratios as a function of the calibration parameter $p$. The x-axis indexes the degree of signal scaling and the y-axis reports the resulting Sharpe ratio. Efficiency peaks at $p=1$ (correct calibration) and declines smoothly as $p$ departs from unity, confirming the cosine-scaling prediction of Proposition~\ref{prop:sharpe_cosine}.
    }
    \label{fig:sharpe_calibration}
  \end{subfigure}
  \caption{\textbf{Calibration and ranking under structural misspecification.} Monte Carlo validation of the theoretical results in Section~\ref{sec:optundermiss}. Panel (a) shows that monotone transformations preserve asset ranking, ensuring feasibility and efficient-set membership (Lemma~\ref{lem:decomposition}). Panel (b) shows that miscalibration alone suffices to attenuate risk-adjusted performance, with Sharpe ratios contracting smoothly as calibration deteriorates (Proposition~\ref{prop:sharpe_cosine}). Together, the panels demonstrate that ranking secures validity of optimization, while calibration determines the attainable position along the efficient frontier.
  }
  \label{fig:ranking_calibration_tradeoff}
\end{figure}

To isolate calibration effects, we generate synthetic return vectors $\mu \in \mathbb{R}^n$ and construct surrogate signals $\tilde{\mu}$ via monotone transformations:
\[
\tilde{\mu}_i = \operatorname{sign}(\mu_i)\,|\mu_i|^p, \quad p \in [0.6,1.4].
\]
These transformations preserve ranking (Lemma~\ref{lem:decomposition}), but distort magnitudes. Panel~(a) of Figure~\ref{fig:ranking_calibration_tradeoff} shows order preservation; Panel~(b) demonstrates that Sharpe efficiency peaks at $p=1$ and contracts smoothly otherwise, confirming Proposition~\ref{prop:sharpe_cosine}. Inefficiency thus stems primarily from calibration errors, not directional failure.

\subsection{Nonlinear Counterexamples and Frontier Geometry}

\begin{figure}[t!]
\centering
\includegraphics[width=0.6\textwidth]{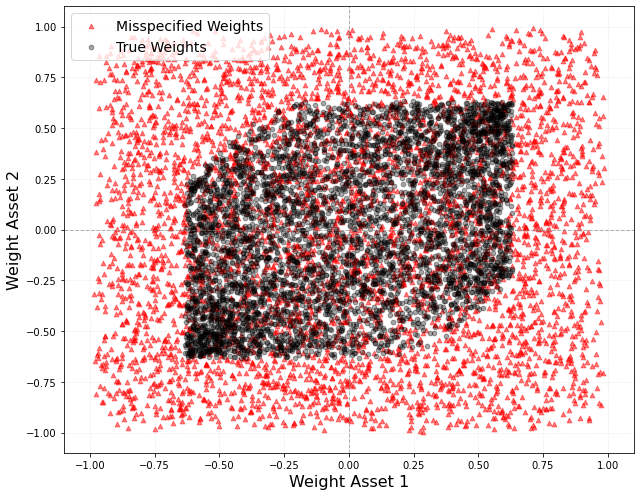}
\caption{\textbf{Directional robustness under nonlinear misspecification.}Scatter of portfolio weights implied by the true return signal (x-axis) versus weights implied by a misspecified predictive signal (y-axis) under a nonlinear confounded data-generating process. Each point corresponds to an asset. While nonlinear misspecification compresses weight magnitudes, the dominant diagonal structure indicates preservation of sign and relative direction. This confirms that nonlinear confounding does not generically induce signal inversion and that directional alignment can persist even when structural correctness fails.
}
\label{fig:weight-scatter-counterexample}
\end{figure}

\begin{figure}[t!]
\centering
\includegraphics[width=0.6\textwidth]{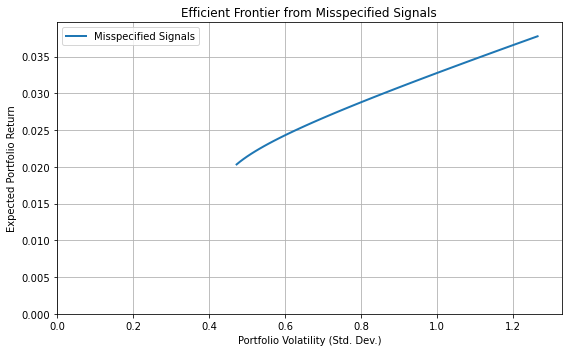}
\caption{\textbf{Efficient frontier viability under misspecified signals.} Mean--variance efficient frontier constructed using misspecified predictive signals. The x-axis reports portfolio volatility and the y-axis expected return implied by the surrogate signal. Despite reduced slope and attainable Sharpe ratios relative to the correctly specified case, the frontier remains smooth, convex, and well defined. This illustrates that structural misspecification degrades efficiency quantitatively through calibration effects but does not collapse optimization geometry.
}
\label{fig:misspecified-frontier}
\end{figure}

We next consider the nonlinear confounded model of Section~\ref{sec:nonlinear}. True and misspecified signals remain directionally aligned (Figure~\ref{fig:weight-scatter-counterexample}), contradicting the claim that omitted variables necessarily induce inversion. Moving to full portfolios, Figure~\ref{fig:misspecified-frontier} shows that efficient frontiers constructed from misspecified signals remain smooth and convex, empirically validating Lemma~\ref{lem:directional-frontier} and Corollary~\ref{cor:frontier_preservation}.

\subsection{Sensitivity to Alignment}

\begin{figure}[t!]
  \centering
  \begin{subfigure}{0.48\textwidth}
    \centering
    \includegraphics[width=\linewidth]{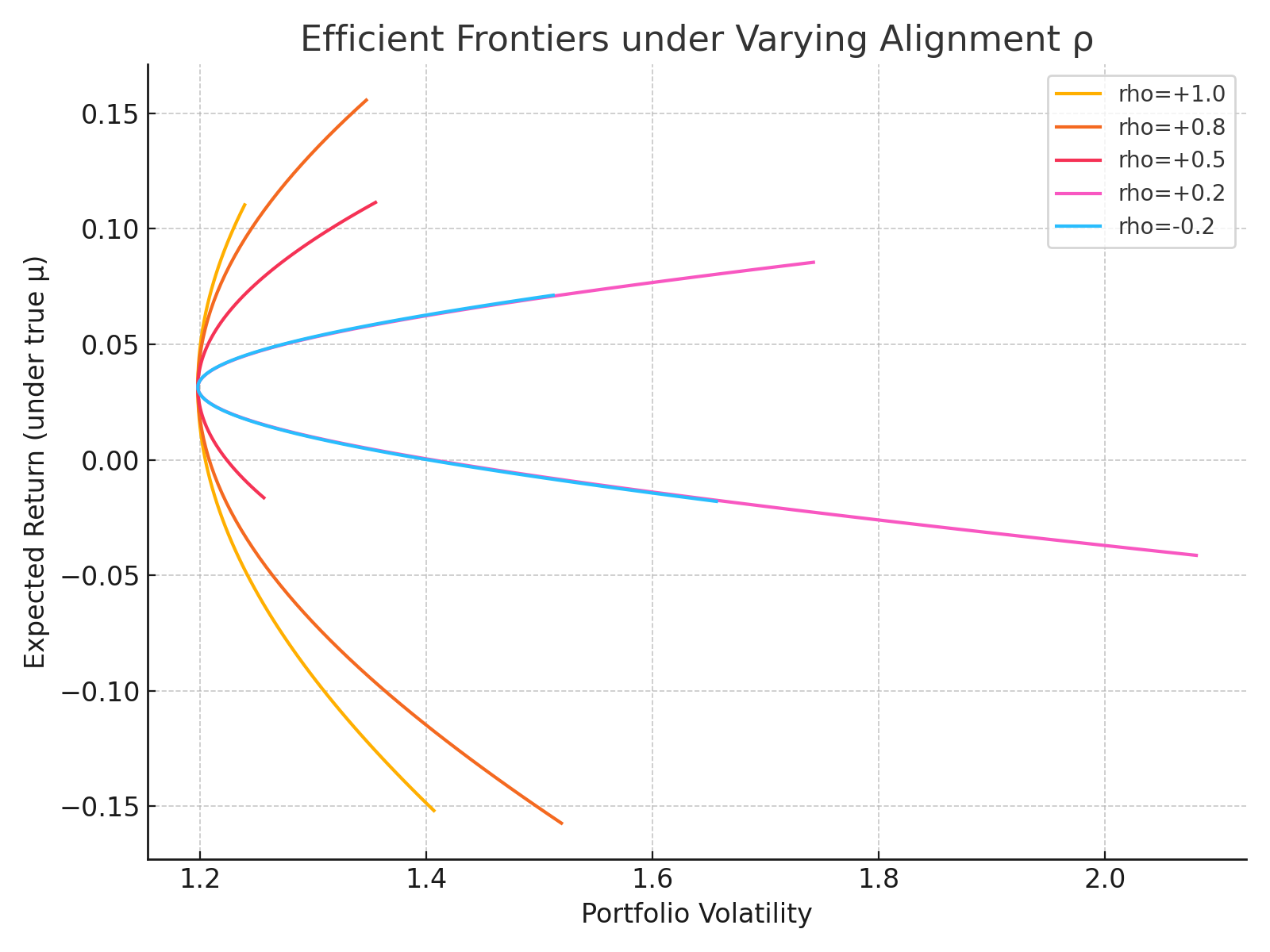}
    \caption{Efficient frontiers constructed from surrogate signals with decreasing cosine alignment $\rho$
    relative to true expected returns. As alignment weakens, attainable risk--return trade-offs contract
    smoothly without loss of convexity or feasibility.
    }
    \label{fig:alignment_frontiers}
  \end{subfigure}
  \hfill
  \begin{subfigure}{0.5\textwidth}
    \centering
    \includegraphics[width=\linewidth]{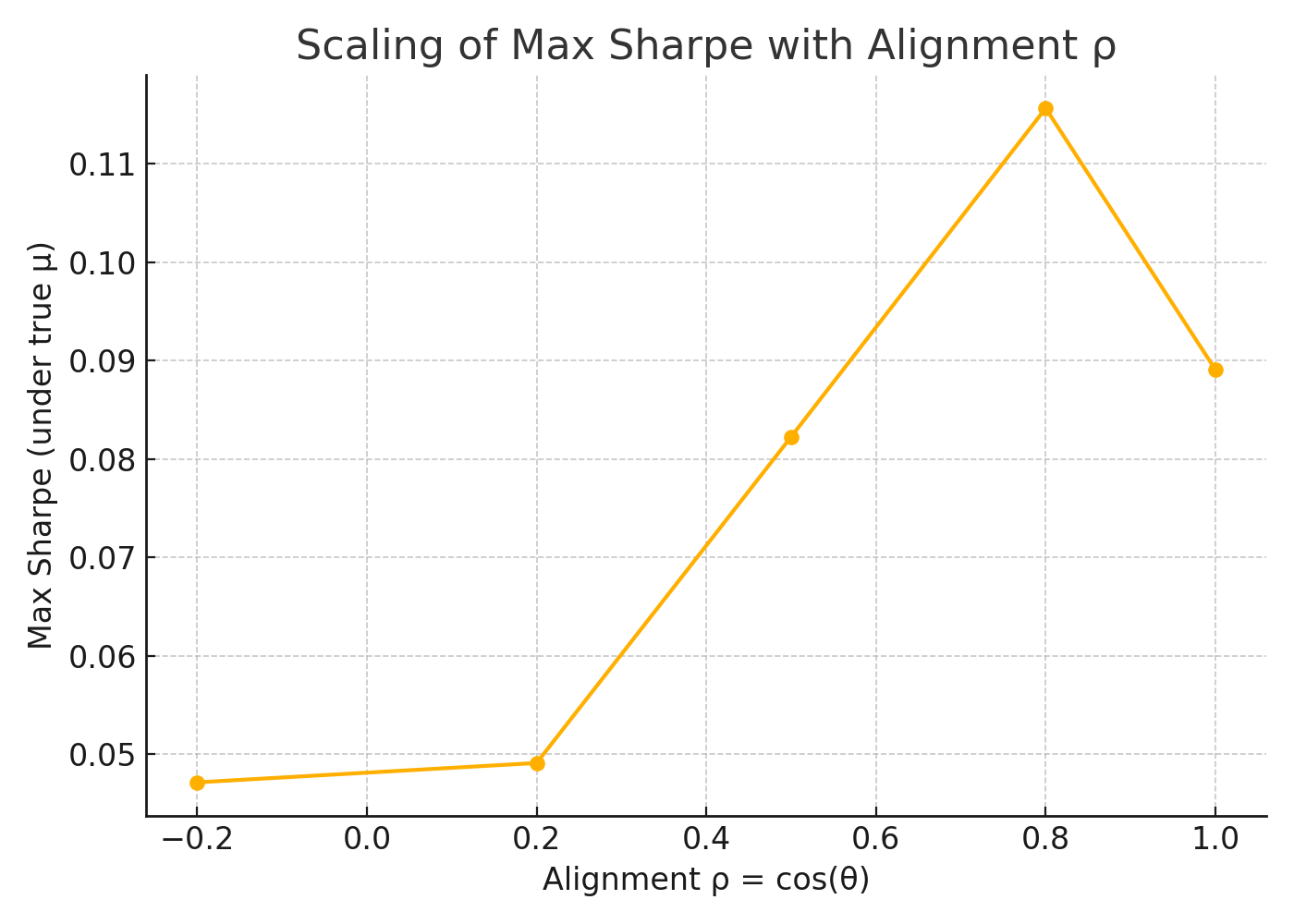}
    \caption{Population Sharpe ratios as a function of cosine alignment $\rho$. Efficiency declines linearly with misalignment, confirming the theoretical scaling law derived in Proposition~\ref{prop:sharpe_alignment_scaling}.
    }
    \label{fig:alignment_sharpe}
  \end{subfigure}
  \caption{\textbf{Sensitivity of Portfolio Efficiency to Directional Alignment.} Panel (a) shows that reduced alignment leads to a proportional contraction of the efficient frontier
  while preserving its geometry. Panel (b) shows that Sharpe ratios scale linearly with alignment. Together, the panels demonstrate that efficiency loss under misspecification arises through smooth geometric contraction rather than frontier collapse.}
  \label{fig:alignment_side_by_side}
\end{figure}

Using surrogate vectors $\tilde{\mu}(\theta)=\cos(\theta)\mu+\sin(\theta)\nu$, we test efficiency as alignment $\rho=\cos(\theta)$ varies. Figure~\ref{fig:alignment_side_by_side} shows (a) frontiers contract proportionally to $\rho$ and (b) Sharpe ratios scale linearly, exactly as predicted by Proposition~\ref{prop:sharpe_alignment_scaling} and Corollary~\ref{cor:frontier_collapse}. These experiments demonstrate that alignment is the critical determinant of efficiency.

\subsection{Empirical Validation on Equity Data}
\label{sec:real_data}

\begin{figure}[ht]
    \centering
    \begin{subfigure}[t]{0.49\textwidth}
        \centering
        \includegraphics[width=\linewidth]{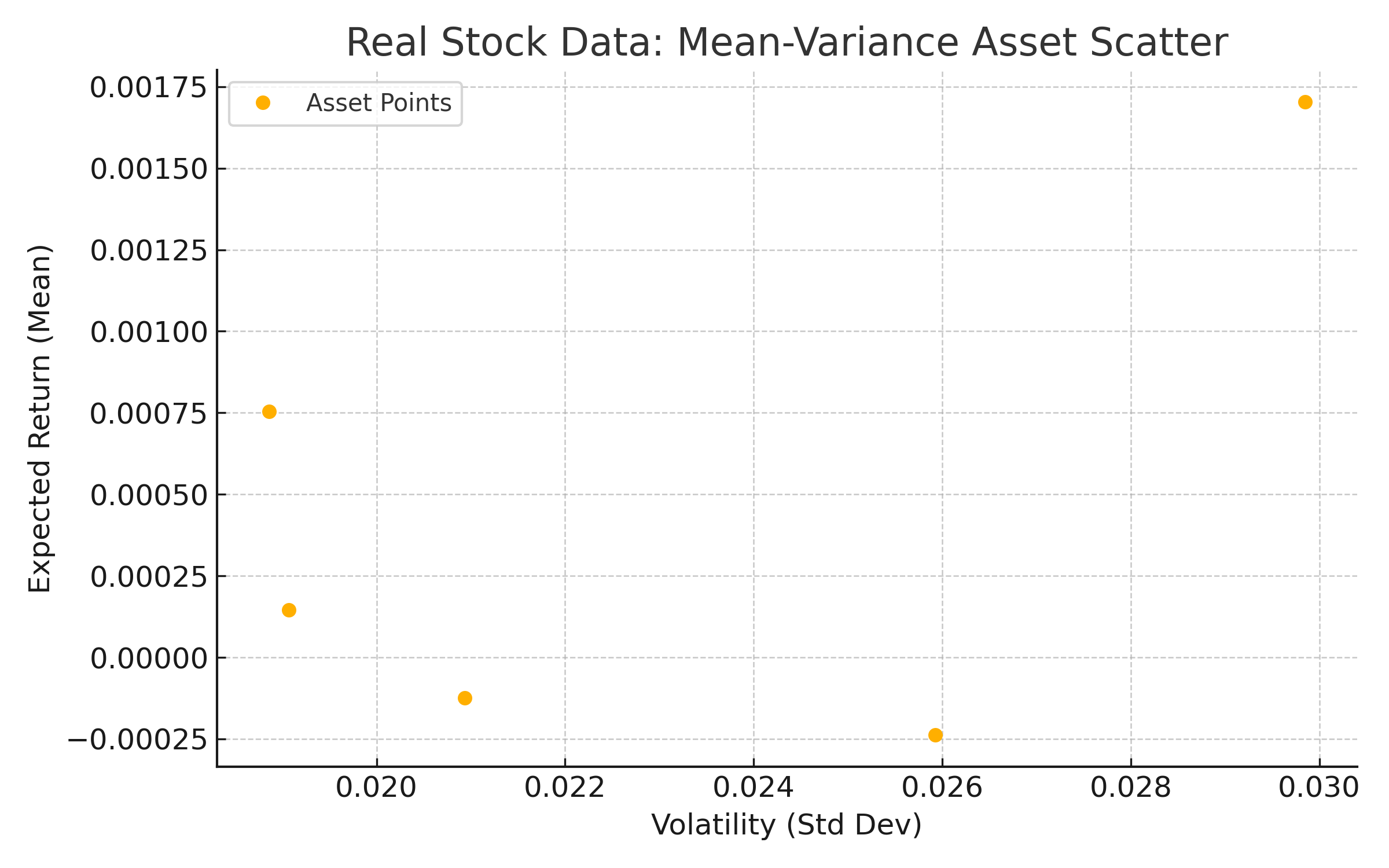}
        \caption{\textbf{Cross-section of S\&P 500 assets.}
        Each point represents a stock, plotted by realized mean return (y-axis) and volatility (x-axis). The dispersion reflects heterogeneous risk--return profiles and provides the empirical geometry on which portfolio diversification operates.
        }
        \label{fig:real_data_scatter}
    \end{subfigure}
    \hfill
    \begin{subfigure}[t]{0.49\textwidth}
        \centering
        \includegraphics[width=\linewidth]{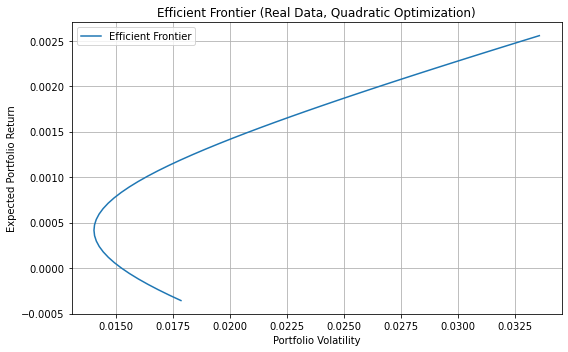}
        \caption{\textbf{Empirical efficient frontier under predictive signals.} Portfolios constructed from predictive (potentially misspecified) signals trace a smooth, convex risk--return frontier. This demonstrates that directional alignment is sufficient to preserve optimization geometry in real data.}
        \label{fig:real_data_frontier}
    \end{subfigure}
    \caption{\textbf{Empirical validation using S\&P 500 equities.} Panel (a) shows the realized cross-sectional risk--return distribution of assets. Panel (b) shows the corresponding empirical efficient frontier obtained via mean--variance optimization using predictive signals. Despite potential misspecification and lack of causal identification, the frontier remains well defined and convex, confirming that directional validity and approximate calibration suffice for actionable portfolio construction in practice.
    }
    \label{fig:real_data_panels}
\end{figure}

Finally, we apply the framework to daily returns of five S\&P 500 stocks over 1000 trading days. Logistic predictors trained on lagged features and volatility measures provide signals $\omega_j$, which feed into a quadratic optimizer. Figure~\ref{fig:real_data_panels} shows that the resulting frontier is smooth and convex despite structural misspecification, confirming that directional validity and approximate calibration suffice for practical viability.

\subsection{Monte Carlo Validation of Bias Attenuation}
\label{sec:approx_cancellation_sim}

\begin{figure}[ht]
    \centering
    \includegraphics[width=0.65\textwidth]{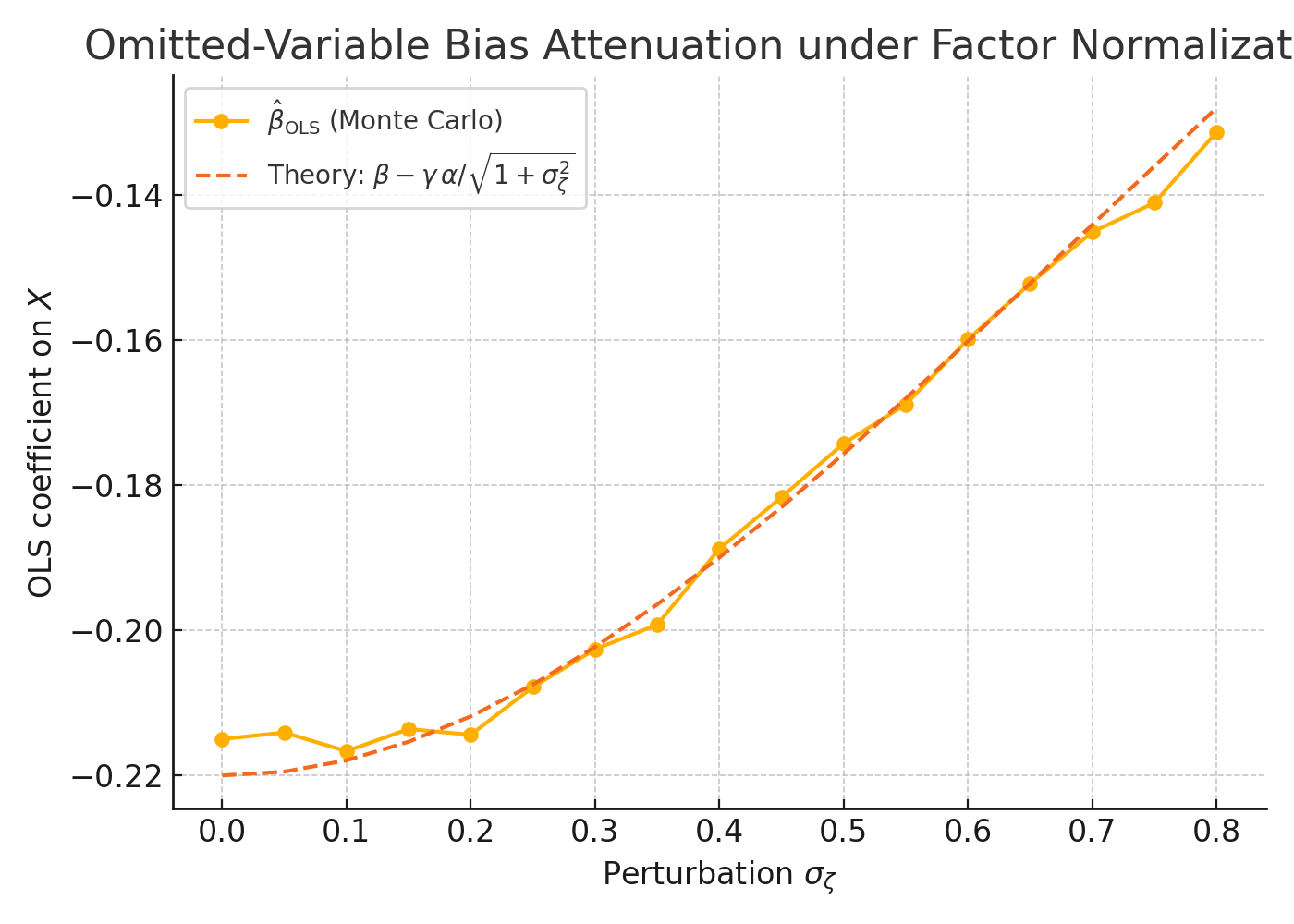}
    \caption{\textbf{Continuous attenuation of omitted-variable bias under confounding noise.} The horizontal axis reports the standard deviation of idiosyncratic confounding noise $\sigma_\zeta$. Dots represent Monte Carlo estimates of the misspecified OLS coefficient $\widehat{\beta}_{\mathrm{OLS}}$, while the dashed line shows the theoretical attenuation curve from Equation~\eqref{eq:attenuation}. Increasing confounding noise attenuates the coefficient smoothly toward the true value, illustrating that omitted-variable bias weakens continuously rather than inducing sign inversion.
    }
    \label{fig:bias_attenuation_combined}
\end{figure}

Varying the confounding noise parameter $\sigma_\zeta$, we regress
$Y=\beta X+\gamma \widetilde{Z}+\varepsilon$ on $X$ alone.
Figure~\ref{fig:bias_attenuation_combined} shows that empirical OLS estimates align closely
with the theoretical attenuation curve from Equation~\eqref{eq:attenuation},
while Table~\ref{tab:bias_values} provides numerical confirmation across noise
levels. Together, these results confirm
Corollary~\ref{cor:ovb_not_failure}: omitted-variable bias degrades predictive
signals smoothly through attenuation rather than inducing catastrophic sign
inversion.

To complement the empirical analysis, Appendix~\ref{app:highD} presents
high-dimensional simulation-based stress tests under controlled misspecification,
designed to isolate the roles of directional alignment, calibration, and
constraints independently of causal structure.

\begin{table}[ht]
\centering
\small
\caption{\textbf{Numerical verification of attenuation law.} Monte Carlo estimates of the misspecified OLS coefficient closely match analytical predictions across confounding-noise levels, confirming the exact bias-decay formula derived in Proposition~\ref{prop:attenuation}.}
\label{tab:bias_values}
\begin{tabular}{r r r}
\toprule
$\sigma_\zeta$ & $\widehat{\beta}_{\mathrm{OLS}}$ & $\beta_{\mathrm{theory}}$ \\
\midrule
0.000 & -0.2203 & -0.2200 \\
0.050 & -0.2187 & -0.2187 \\
0.100 & -0.2163 & -0.2163 \\
0.200 & -0.2099 & -0.2099 \\
0.300 & -0.2020 & -0.2020 \\
0.400 & -0.1929 & -0.1929 \\
0.500 & -0.1828 & -0.1829 \\
0.600 & -0.1720 & -0.1720 \\
0.700 & -0.1605 & -0.1606 \\
0.800 & -0.1486 & -0.1486 \\
\bottomrule
\end{tabular}
\end{table}

\section{Discussion}
\label{sec:discussion}

This paper revisits the claim that causal factor models are a prerequisite for
portfolio optimization. Through theoretical results, simulation counterexamples,
and empirical validation, we show that this assertion does not hold within the static mean--variance and closely related quadratic portfolio optimization frameworks analyzed in this paper. Even when structurally misspecified, predictive models can produce
signals that remain directionally aligned with true returns and are sufficient to
sustain smooth, convex efficient frontiers.

\subsection{On the Claim That Misspecification Necessarily Implies Signal Inversion}

A recurring assertion in recent critiques is that structural misspecification, including omitted variables and nonlinear confounding, inevitably produces inverted predictive signals and collapses the efficient frontier. Our results directly contradict this claim within the static mean--variance portfolio optimization framework analyzed in this paper. Sections~\ref{sec:structural_cancellation} and~\ref{sec:nonlinear} provide explicit counterexamples showing that omitted-variable bias can induce attenuation without inversion, preserving directional informativeness under moderate departures from the true data-generating process.

Section~\ref{sec:approx_cancellation} further demonstrates that signal degradation under misspecification is not knife-edge. Both analytically and empirically, omitted-variable bias attenuates continuously as dependence weakens (Figure~\ref{fig:bias_attenuation_combined} and Table~\ref{tab:bias_values}). This continuity result highlights a precise notion of robustness: optimization performance degrades smoothly rather than collapsing abruptly.

\subsection{On the Claim That Inefficiency Requires Signal Inversion}

A second, closely related claim is that inefficiency in portfolio optimization arises only through sign inversion. Our analysis clarifies why this is incorrect within the static mean--variance framework. We explicitly separate directionality, ranking, and calibration. Theorem~\ref{thm:alignment_efficiency}, Lemma~\ref{lem:ranking_not_sizing}, and Corollary~\ref{cor:calibration_necessary} establish that directional correctness and ranking are necessary but not sufficient for efficiency: miscalibration alone is sufficient to attenuate realized returns and Sharpe ratios.

This refines the interpretation of the sort-based framework of
\citet{almgren2005sorts}. While correct ranking ensures efficient-set membership, Proposition~\ref{prop:sharpe_cosine} and Lemma~\ref{lem:decomposition} show that calibration governs quantitative performance within that set. Figures~\ref{fig:ranking_calibration_tradeoff}
and~\ref{fig:alignment_side_by_side} confirm empirically that inefficiency most often arises through magnitude distortion rather than sign failure.

\subsection{On the Robustness of Optimization Geometry}

Beyond signal properties, we examine whether optimization geometry itself is
fragile to misspecification. Lemma~\ref{lem:directional-frontier} and
Proposition~\ref{prop:sharpe_alignment_scaling} show that, as long as surrogate
signals remain directionally aligned with true expected returns, the efficient
frontier remains smooth, convex, and non-degenerate within the static mean--variance optimization problem with a covariance risk model. Structural misspecification does not collapse optimization geometry.

In this context, robustness refers to the preservation of the mean--variance efficient frontier geometry under misspecified inputs, rather than invariance of optimal weights or realized returns. The static mean--variance problem remains well posed under a wide class of misspecifications: feasibility and convexity are preserved, while investment efficiency degrades smoothly through miscalibration rather than failing discontinuously.

Our empirical study on S\&P~500 equities (Figure~\ref{fig:real_data_panels}) confirms that these theoretical properties persist in practice. Even without structural correctness, directional validity combined with approximate calibration suffices for coherent and actionable portfolio design.

\subsection{On the Role of Causality Beyond Optimization}
\label{sec:practical_implications}

The results of this paper do not argue against the use of causal inference in finance. Causal models remain essential for tasks such as structural attribution, regime analysis, stress testing, and counterfactual reasoning. The contribution of this work is instead to clarify a boundary condition: causal identifiability is not a prerequisite for portfolio efficiency within the static mean--variance and closely related quadratic portfolio optimization frameworks, provided that predictive signals satisfy empirically verifiable geometric conditions.

From an operational perspective, the relevant object for portfolio construction is not causal structure itself, but the empirical geometry induced by predictive signals in interaction with the risk model. Portfolio optimization remains viable whenever surrogate return vectors preserve positive directional alignment and stable ranking relative to realized returns. Structural misspecification primarily affects performance through miscalibration of signal magnitudes rather
than systematic sign inversion.

This distinction has direct practical consequences. Predictive models (whether causally motivated or purely associational) can be evaluated for portfolio suitability using observable diagnostics, including out-of-sample directional alignment, ranking stability across rolling windows, and sensitivity of portfolio performance to signal rescaling. When these conditions are satisfied, optimization remains well posed, and calibration determines the attainable position along the efficient frontier.

Causal information becomes relevant for portfolio optimization only insofar as it informs the construction or stabilization of the diversification geometry itself, rather than as a requirement imposed on return prediction. Within any such geometry, investment efficiency continues to be governed by alignment, ranking, and calibration. This perspective provides a principled role for causality while preserving a clear separation between explanatory objectives and optimization
feasibility.

\section{Conclusion and Future Work}
\label{sec:conclusion}

We have re-evaluated the claim that causal factor models are necessary for portfolio efficiency within the static mean--variance and closely related quadratic portfolio optimization frameworks. Through formal analysis, controlled simulations, and empirical evidence, we show that structural misspecification, including omitted variables and nonlinear confounding, does not inherently invalidate portfolio optimization. Predictive models without causal identifiability can nevertheless yield directionally informative signals that sustain smooth and convex efficient frontiers.

Our contributions can be summarized as follows. First, we disentangle directionality, ranking, and calibration, establishing that directional correctness is necessary but not sufficient for portfolio efficiency when signal magnitudes are miscalibrated. Second, we show that structurally misspecified yet directionally aligned signals
preserve optimization geometry, extending the portfolios-from-sorts framework of \citet{almgren2005sorts} by explicitly accounting for calibration effects within the efficient set. Third, we demonstrate analytically and empirically that misspecification induces smooth attenuation rather than abrupt collapse of efficiency. Fourth, we validate these results with real market data, confirming that
predictive signals can support coherent and actionable frontiers even in the absence of strict causal grounding.

These findings caution against overstating the necessity of causal inference in portfolio design. While causal models remain indispensable for attribution, regime analysis, and stress testing
\citep{09641e90-d989-3dd6-88ae-92e35971022e, Wilcox2014HierarchicalCI}, causal
identifiability is not intrinsic to the objective of portfolio efficiency within the static mean--variance and closely related quadratic optimization frameworks analyzed here. Optimization performance is governed by the geometric relationship between predictive signals and realized returns, specifically directional alignment, ranking stability, and calibration, rather than by structural correctness per se. More broadly, the results clarify a recurring misconception in the literature: improvements in causal validity or predictive accuracy at the asset level do not, by themselves, determine investment efficiency. Portfolio optimization is fundamentally a diversification problem defined relative to a chosen risk--return geometry. For example, causal information can be incorporated into portfolio construction by shaping the diversification geometry itself, such as through sensitivity-based representations of common causal drivers \citep{RODRIGUEZDOMINGUEZ2023100447}. In such approaches, causality informs the structure of the risk--return space, while investment efficiency remains governed by the geometric properties of the resulting optimization problem.

When causality is incorporated into portfolio construction, it operates by shaping or stabilizing the diversification geometry itself rather than as a requirement imposed on return prediction. Even in such settings, investment efficiency remains governed by alignment, ranking, and calibration within the induced space. Future work may extend these results in several directions, including integrating causal diagnostics into predictive pipelines without conflating them with efficiency criteria, examining the geometric sufficiency conditions under alternative risk functionals and non-quadratic objectives, and studying misspecification-resilient optimization in higher-dimensional and nonlinear settings.

\appendix
\section{Simulation Protocols and Code}
\label{appendix:simulation}

All simulations were implemented in Python using NumPy, Matplotlib, and
scikit-learn where necessary. Random seeds were fixed for reproducibility. The protocols described in this appendix support the numerical results in
Section~\ref{sec:results} and the
high-dimensional stress tests reported in Appendix~\ref{app:highD}. Across experiments:

\begin{itemize}
  \item Synthetic features \( X_1, X_2 \sim \mathcal{N}(0,1) \) were generated independently for \( n = 1000 \) observations unless otherwise stated.
  \item Confounders \( Z \) were constructed as linear combinations of \( X_1, X_2 \) with Gaussian noise \( \eta \sim \mathcal{N}(0,1) \).
  \item Nonlinear transformations such as \( \tanh(\cdot) \) and \( \sin(\cdot) \) were used to define signal functions.
  \item True outcomes \( Y \sim \text{Bernoulli}(\sigma(f(X) + g(Z))) \) were generated given these transformations.
  \item Logistic regressions were trained using only \( X \), omitting \( Z \), to simulate structural misspecification.
  \item Portfolio weights were constructed as \( \omega = 2 \hat{p}(X) - 1 \), where \( \hat{p} \) is the predicted probability.
  \item Optimization used empirical mean and covariance matrices to solve the classical Markowitz problem with quadratic programming (via CVXPY).
\end{itemize}

Monte Carlo experiments (Section~\ref{sec:approx_cancellation_sim}) used \( n = 200{,}000 \) samples and varied \( \sigma_\zeta \in [0, 0.8] \). OLS regressions were computed with NumPy’s least squares solver. All empirical results were averaged over 10 repetitions.

\textit{Simulation return mapping.}
In simulation-based experiments that employ probabilistic classifiers
(Sections~\ref{sec:nonlinear} and~\ref{sec:approx_cancellation}),
predictive models output class probabilities
$p(x)=\mathbb{P}(Y=1\mid x)$, which are mapped to realized returns via the affine
transformation $r(x)=a\,(2p(x)-1)+\varepsilon$, with $a>0$.
This mapping preserves directional alignment, ranking, and cosine geometry with
respect to the true signal, ensuring consistency with the linear
mean--variance framework analyzed in the theory.
All results in these experiments depend only on directional alignment and
calibration, not on the specific affine scaling.

\subsection{Code for Calibration–Ranking Experiments}
\label{appendix:rankingcalib}

\begin{verbatim}
import os
import numpy as np
import matplotlib.pyplot as plt

np.random.seed(42)
os.makedirs("figures", exist_ok=True)

def make_spd_cov(n, n_factors=3, idio_scale=0.2):
    F = np.random.randn(n, n_factors)
    Lambda = np.diag(np.random.uniform(0.6, 1.4, size=n_factors))
    Sigma_f = F @ Lambda @ F.T
    D = np.diag(np.random.uniform(idio_scale, 2*idio_scale, size=n))
    V = Sigma_f + D
    V = 0.5*(V + V.T)
    w, Q = np.linalg.eigh(V)
    w = np.maximum(w, 1e-8)
    V = (Q * w) @ Q.T
    Vinv = (Q * (1.0/w)) @ Q.T
    return V, Vinv

def vminner(u, v, Vinv): return float(u.T @ Vinv @ v)
def vnorm(u, Vinv): return np.sqrt(max(vminner(u,u,Vinv), 1e-18))
def tangency_direction(signal, Vinv): return Vinv @ signal
def mean_vol_sharpe(w, mu, V):
    mean = float(mu.T @ w)
    vol = float(np.sqrt(max(w.T @ V @ w, 1e-18)))
    return mean, vol, mean/vol if vol>0 else np.nan

def generate_mu(n=100, scale=0.25):
    mu = np.random.randn(n)
    return mu / (np.linalg.norm(mu)+1e-12) * scale

def monotone_transforms(mu, powers):
    return [np.sign(mu) * (np.abs(mu)**p) for p in powers]

def main():
    n, powers = 120, (0.6, 0.8, 1.0, 1.2, 1.4)
    V, Vinv = make_spd_cov(n)
    mu = generate_mu(n=n, scale=0.25)

    tildes_raw = monotone_transforms(mu, powers)
    mu_norm = vnorm(mu, Vinv)
    tildes = [t*(mu_norm/(vnorm(t,Vinv)+1e-12)) for t in tildes_raw]

    # (a) Scatter
    plt.figure()
    for k in [0,2,4]:
        p = powers[k]
        plt.scatter(mu, tildes[k], s=12, alpha=0.55, label=f"p={p:.1f}")
    plt.axhline(0,color="black",lw=0.6,alpha=0.4)
    plt.axvline(0,color="black",lw=0.6,alpha=0.4)
    plt.xlabel(r"$\mu$")
    plt.ylabel(r"$\tilde{\mu}$")
    plt.title("Ranking preserved under monotone transforms")
    plt.legend(); plt.grid(True,ls="--",alpha=0.35)
    plt.tight_layout()
    plt.savefig("figures/ranking_preserved.png", dpi=220)

    # (b) Sharpe ratios
    rel_sharpes, sharpe_true = [], None
    for p, t in zip(powers,tildes):
        _,_,S_mu = mean_vol_sharpe(tangency_direction(mu,Vinv), mu, V)
        _,_,S_t  = mean_vol_sharpe(tangency_direction(t,Vinv), mu, V)
        if sharpe_true is None: sharpe_true = S_mu
        rel_sharpes.append(S_t/(S_mu+1e-12))

    plt.figure()
    plt.plot(powers, rel_sharpes, marker="o")
    plt.xlabel("Exponent $p$")
    plt.ylabel("Relative Sharpe")
    plt.title("Sharpe efficiency vs. calibration")
    plt.grid(True,ls="--",alpha=0.35)
    plt.tight_layout()
    plt.savefig("figures/sharpe_calibration.png", dpi=220)

if __name__=="__main__":
    main()
\end{verbatim}

\subsection{Simulated Experiment for Misspecified Signal Frontier}
\begin{verbatim}
import numpy as np
import matplotlib.pyplot as plt
import cvxpy as cp

# Simulation parameters
np.random.seed(42)
n_assets = 5
n_obs = 1000

# Features and confounder
X = np.random.normal(0, 1, size=(n_obs, n_assets))
Z = 0.7 * X[:, 0] + 0.3 * X[:, 1] + np.random.normal(0, 1, size=n_obs)
logit_true = np.tanh(X) + 0.5 * np.sin(Z)[:, None]
p_true = 1 / (1 + np.exp(-logit_true))

# True returns
returns = (2 * p_true - 1) * 2.0 + np.random.normal(0, 0.05, (n_obs, n_assets))

# Misspecified logistic models per asset
betas = np.random.normal(1.0, 0.5, size=(n_assets, n_assets))
logits = X @ betas.T
p_pred = 1 / (1 + np.exp(-logits))
signals = 2 * p_pred - 1

# Empirical moments
mu_hat = signals.mean(axis=0)
Sigma = np.cov(returns, rowvar=False)

# Target returns and frontier
target_returns = np.linspace(mu_hat.min() * 1.5, mu_hat.max() * 1.5, 50)
efficient_risks = []
efficient_returns = []

for R in target_returns:
    w = cp.Variable(n_assets)
    prob = cp.Problem(cp.Minimize(cp.quad_form(w, Sigma)),
                      [cp.sum(w) == 1, mu_hat @ w == R])
    prob.solve()
    if w.value is not None:
        vol = np.sqrt(w.value @ Sigma @ w.value)
        efficient_returns.append(R)
        efficient_risks.append(vol)

# Plot
plt.figure(figsize=(8, 5))
plt.plot(efficient_risks, efficient_returns,
         label='Misspecified Signals', color='tab:blue')
plt.xlabel("Portfolio Volatility")
plt.ylabel("Expected Portfolio Return")
plt.title("Efficient Frontier from Misspecified Signals")
plt.grid(True)
plt.legend()
plt.tight_layout()
plt.savefig("efficient_frontier_qp_misspecified_scaled.png", dpi=300)
plt.show()
\end{verbatim}

\subsection{Efficient Frontier from Real Financial Data}
\label{appendixfrontier}
\begin{verbatim}
import numpy as np
import pandas as pd
import matplotlib.pyplot as plt
import cvxpy as cp

# Load Excel file
file_path = "constituents_SPX_2008_Full.xlsx"
df = pd.read_excel(file_path)

# Date processing and returns
df['Date'] = pd.to_datetime(df['Date'], format='%Y%m%d')
df.set_index('Date', inplace=True)
returns = df.pct_change().replace([np.inf, -np.inf], np.nan).dropna()

# Subset selection: 5 assets over 1000 days
subset_returns = returns.iloc[:1000, :5]
mu_hat = subset_returns.mean().values
Sigma = subset_returns.cov().values
n_assets = len(mu_hat)

# Efficient frontier computation
target_returns = np.linspace(mu_hat.min() * 1.5, mu_hat.max() * 1.5, 50)
efficient_risks = []
efficient_means = []

for R in target_returns:
    w = cp.Variable(n_assets)
    constraints = [cp.sum(w) == 1, mu_hat @ w == R]
    problem = cp.Problem(cp.Minimize(cp.quad_form(w, Sigma)), constraints)
    problem.solve()

    if w.value is not None:
        port_return = mu_hat @ w.value
        port_vol = np.sqrt(w.value @ Sigma @ w.value)
        efficient_means.append(port_return)
        efficient_risks.append(port_vol)

# Plotting
plt.figure(figsize=(8, 5))
plt.plot(efficient_risks, efficient_means, label='Efficient Frontier', color='tab:blue')
plt.xlabel("Portfolio Volatility")
plt.ylabel("Expected Portfolio Return")
plt.title("Efficient Frontier (Real Data, Quadratic Optimization)")
plt.grid(True)
plt.legend()
plt.tight_layout()
plt.show()
\end{verbatim}

\subsection{Code for Bias Attenuation Simulation}

\begin{verbatim}
import numpy as np
import matplotlib.pyplot as plt

# Parameters
alpha = 0.6
beta = 0.2
gamma = 0.7
n = 200_000
sigma_zeta_vals = np.linspace(0.0, 0.8, 17)

empirical_beta = []
theoretical_beta = []

# Loop over different confounding noise levels
for sigma_zeta in sigma_zeta_vals:
    X = np.random.normal(0, 1, size=n)
    eta = np.random.normal(0, np.sqrt(1 - alpha**2), size=n)
    zeta = np.random.normal(0, sigma_zeta, size=n)

    Z = -alpha * X + eta + zeta
    # sigma_eta^2 fixed to 1 and absorbed into baseline normalization
    Z_tilde = Z / np.sqrt(1 + sigma_zeta**2)

    epsilon = np.random.normal(0, 1, size=n)
    Y = beta * X + gamma * Z_tilde + epsilon

    # OLS regression of Y on X alone
    X_centered = X - X.mean()
    beta_hat = np.dot(X_centered, Y - Y.mean()) / np.dot(X_centered, X_centered)
    empirical_beta.append(beta_hat)

    # Theoretical OVB formula
    beta_theory = beta - gamma * alpha / np.sqrt(1 + sigma_zeta**2)
    theoretical_beta.append(beta_theory)

# Plot results
plt.figure(figsize=(8, 5))
plt.plot(sigma_zeta_vals, empirical_beta, 'o', label='Monte Carlo')
plt.plot(sigma_zeta_vals, theoretical_beta, '--', label='Theory')
plt.xlabel(r"$\sigma_\zeta$")
plt.ylabel(r"$\hat{\beta}_{\mathrm{OLS}}$")
plt.title("Bias Attenuation under Approximate Cancellation")
plt.legend()
plt.grid(True)
plt.tight_layout()
plt.savefig("appendix_E_bias_attenuation.png", dpi=300)
plt.show()
\end{verbatim}

\section{Refutation of Structural Necessity Claims}
\label{appendix:refutation}

This appendix corresponds to the refutation roadmap outlined in the Introduction and summarized in Table~\ref{tab:refutation-summary}, providing formal and empirical resolution of the necessity claims discussed in the main text within the class of static portfolio optimization frameworks considered in this paper.

Several recent critiques of predictive approaches to portfolio optimization 
\citep{lopez2024case, lopez2025causal} have advanced strong structural arguments suggesting that causal factor models are not only desirable but strictly necessary for efficiency. These arguments rest on claims such as: (i) misspecification inevitably produces sign inversion; 
(ii) sign inversion is the sole driver of inefficiency; 
(iii) sign agreement guarantees efficiency; and 
(iv) coherent efficient frontiers cannot exist without fully specified causal models. 

The framework developed in this paper allows us to test these propositions formally. 
Across theory, controlled simulations, and empirical analysis, we show that these claims do not hold once more general conditions are considered. 
Misspecified models often preserve directional alignment, inefficiency frequently arises from calibration errors rather than inversion, and efficient frontiers remain convex provided that signals retain positive alignment with expected returns. 
Taken together, these results demonstrate that the essential determinants of viability are directionality and calibration, not structural fidelity. 
Misspecification can attenuate performance, but it does not by itself collapse optimization geometry or invalidate predictive frontiers.

Importantly, these conclusions do not assert that causal structure is irrelevant, but rather that causal identifiability is not a necessary condition for portfolio efficiency within the optimization settings analyzed here.

Table~\ref{tab:refutation-summary} summarizes the principal claims, their logical form, and the corresponding refutations established in this paper. 
The recurring message is clear: predictive signals need not be causally identified to support coherent portfolio construction, and the mechanisms of inefficiency are more nuanced than previously asserted. 

\begin{table}[H]
\centering
\small
\renewcommand{\arraystretch}{1.3}
\begin{tabular}{|p{4.8cm}|p{4.8cm}|p{5.0cm}|}
\hline
\textbf{Claim from Prior Work} & \textbf{Logical Form} & \textbf{Refutation in This Paper} \\
\hline

Misspecification necessarily causes sign inversion 
& \(\text{Misspecification} \Rightarrow \text{Sign Inversion}\) 
& Counterexamples in Sections~\ref{sec:structural_cancellation} and~\ref{sec:nonlinear} show that alignment can persist under omitted-variable bias. Figures~\ref{fig:weights-cancellation} and~\ref{fig:weight-scatter-counterexample} confirm empirically that inversion is not inevitable. \\ \hline

Sign inversion is required for inefficiency 
& \(\text{Inefficiency} \Leftarrow \text{Sign Inversion}\) 
& Theorem~\ref{thm:alignment_efficiency} proves that inefficiency arises from miscalibration even without inversion. Lemma~\ref{lem:ranking_not_sizing} and Corollary~\ref{cor:calibration_necessary} show that correct ranking with poor sizing remains inefficient. \\ \hline

Sign agreement guarantees efficiency 
& \(\text{Sign Agreement} \Rightarrow \text{Efficiency}\) 
& Refuted by Theorem~\ref{thm:alignment_efficiency}: aligned signals with distorted magnitudes yield attenuated Sharpe ratios. Efficiency requires both alignment and calibration. \\ \hline

Misspecification invalidates mean--variance optimization 
& \(\text{Misspecification} \Rightarrow \text{No Frontier}\) 
& Lemma~\ref{lem:directional-frontier} and Corollary~\ref{cor:frontier_preservation} prove that positive alignment suffices for smooth, convex frontiers. Section~\ref{sec:results} (Figures~\ref{fig:misspecified-frontier}, \ref{fig:alignment_side_by_side}) confirms this empirically. \\ \hline

Only causal factor models support efficiency 
& \(\neg \text{Causal} \Rightarrow \neg \text{Efficient}\) 
& Refuted theoretically and empirically: Section~\ref{sec:real_data} shows that associational signals from lagged, non-causal features yield viable frontiers. Section~\ref{sec:3} formalizes that efficiency depends on directional informativeness, not causal identification. \\ \hline

Misspecified signals produce anti-aligned weights 
& \( \omega^{\text{pred}} \approx -\omega^{\text{true}} \) 
& No general basis. Figure~\ref{fig:weight-scatter-counterexample} shows no clustering along the negative diagonal. Section~\ref{sec:approx_cancellation} and Monte Carlo experiments (Section~\ref{sec:approx_cancellation_sim}) demonstrate smooth attenuation, not inversion. \\ \hline
\end{tabular}
\caption{Structural claims from \citep{lopez2024case, lopez2025causal} and their refutations. Across cases, inefficiency is shown to arise from miscalibration rather than inevitable inversion, and viable frontiers persist under positive alignment.}
\label{tab:refutation-summary}
\end{table}

\section{Correction of Variance Normalization in Factor Models}
\label{appendix:variance}

In Section~\ref{sec:prelim}, we identified a normalization discrepancy in \citep{lopez2024case}, where the reported variance of \( F_2 \) is inconsistent with standard factor modeling assumptions. We now provide the corrected derivation and its implications for bias interpretation.

\subsection{Setup}

Let:
\[
F_2 = \delta Z + \eta, \quad Z \sim \mathcal{N}(0,1), \quad \eta \sim \mathcal{N}(0,1-\delta^2), \quad Z \perp \eta.
\]
Then \( \mathrm{Var}(F_2) = \delta^2 + (1-\delta^2) = 1 \), and \( \mathrm{Cov}(Z,F_2)=\delta \), maintaining unit variance and interpretable loadings.

\subsection{Comparison}

\begin{table}[h]
\centering
\small
\renewcommand{\arraystretch}{1.3}
\begin{tabular}{|p{5.3cm}|p{4.0cm}|p{5.0cm}|}
\hline
\textbf{Assumption on $\mathrm{Var}(F_2)$} & \textbf{Estimated Loading $\hat{\beta}_n$} & \textbf{Bias Characterization} \\
\hline
Correct: $\mathrm{Var}(F_2) = 1$ 
& $\hat{\beta}_n = \beta_n + \gamma_n \delta$ 
& Linear in $\delta$; interpretable directional bias \\
\hline
\citep{lopez2024case}: $\mathrm{Var}(F_2) = 1+\delta^2$ 
& $\hat{\beta}_n = \dfrac{\beta_n + \gamma_n \delta}{1+\delta^2}$ 
& Nonlinear attenuation; no longer proportionate to confounding strength \\
\hline
\end{tabular}
\caption{Impact of variance normalization on estimated factor loadings and omitted-variable bias interpretation.}
\label{tab:variance-comparison}
\end{table}

Despite functional differences, both formulations confirm that confounders induce signal bias. However, only the corrected form maintains scale-consistent estimation and analytically tractable bias propagation.

\section{Full Monte Carlo Results for Approximate Cancellation}
\label{fulltable}

This appendix complements Section~\ref{sec:approx_cancellation_sim} by reporting the full numerical table of simulation results for omitted-variable bias under increasing structural noise. See Appendix~\ref{appendix:simulation} for implementation.

\begin{table}[H]
\centering
\small
\renewcommand{\arraystretch}{1.15}
\begin{tabular}{r r r r r}
\toprule
$\sigma_\zeta$ & $\hat{\beta}_{\mathrm{OLS}}$ (MC) & $\beta_{\mathrm{theory}}$ & Bias (MC) & Bias (Theory) \\
\midrule
0.000 & -0.220295 & -0.220000 & -0.420295 & -0.420000 \\
0.050 & -0.218724 & -0.218690 & -0.418724 & -0.418690 \\
0.100 & -0.216333 & -0.216330 & -0.416333 & -0.416330 \\
0.150 & -0.213370 & -0.213374 & -0.413370 & -0.413374 \\
0.200 & -0.209934 & -0.209951 & -0.409934 & -0.409951 \\
0.250 & -0.206124 & -0.206150 & -0.406124 & -0.406150 \\
0.300 & -0.201989 & -0.202021 & -0.401989 & -0.402021 \\
0.350 & -0.197564 & -0.197600 & -0.397564 & -0.397600 \\
0.400 & -0.192882 & -0.192920 & -0.392882 & -0.392920 \\
0.450 & -0.187965 & -0.188005 & -0.387965 & -0.388005 \\
0.500 & -0.182833 & -0.182874 & -0.382833 & -0.382874 \\
0.550 & -0.177504 & -0.177544 & -0.377504 & -0.377544 \\
0.600 & -0.172000 & -0.172039 & -0.372000 & -0.372039 \\
0.650 & -0.166337 & -0.166373 & -0.366337 & -0.366373 \\
0.700 & -0.160533 & -0.160566 & -0.360533 & -0.360566 \\
0.750 & -0.154606 & -0.154636 & -0.354606 & -0.354636 \\
0.800 & -0.148573 & -0.148600 & -0.348573 & -0.348600 \\
\bottomrule
\end{tabular}
\caption{Monte Carlo and theoretical results for omitted-variable bias under approximate cancellation. Simulation size: $n = 200{,}000$. Parameters: $\alpha=0.6$, $\beta=0.2$, $\gamma=0.7$.}
\label{tab:ovb_bias_table}
\end{table}

\section{High-Dimensional Stress Tests under Controlled Misspecification}
\label{app:highD}

This appendix provides the additional robustness analyses. It reports stress tests addressing high dimensionality,
covariance regularization, and constraint-induced attenuation. The experiments
cover regimes with $N \gg T$, singular and shrinkage-estimated covariance matrices,
and binding portfolio constraints. Across all settings, the attenuation behavior
derived in the main text persists without qualitative change: portfolio efficiency
degrades smoothly rather than exhibiting threshold effects or instability.

All results in this appendix are obtained from synthetic data to isolate geometric
and informational effects from market-specific structure. Simulations are fully
replicable, implemented in Python with fixed pseudo-random seeds, and involve no
ex post parameter tuning. We consider asset universes with $N \in \{300,500,800,1000\}$. Returns follow a
factor-based covariance structure $\Sigma = BB^\top + D$, with $B$ and $D$
generated independently. Expected returns $\mu$ are drawn independently of
$\Sigma$. Predictive signals are constructed as directional surrogates
\[
\tilde{\mu}(\theta) = \cos(\theta)\mu + \sin(\theta)\nu,
\]
where $\nu$ is orthogonal to $\mu$. Varying $\theta \in [0,\pi]$ controls alignment
without altering marginal distributions. Portfolio weights are computed under
mean--variance geometry, and performance is evaluated using population Sharpe
ratios under the true moments $(\mu,\Sigma)$.

\begin{figure}[t]
\centering
\includegraphics[width=0.6\linewidth]{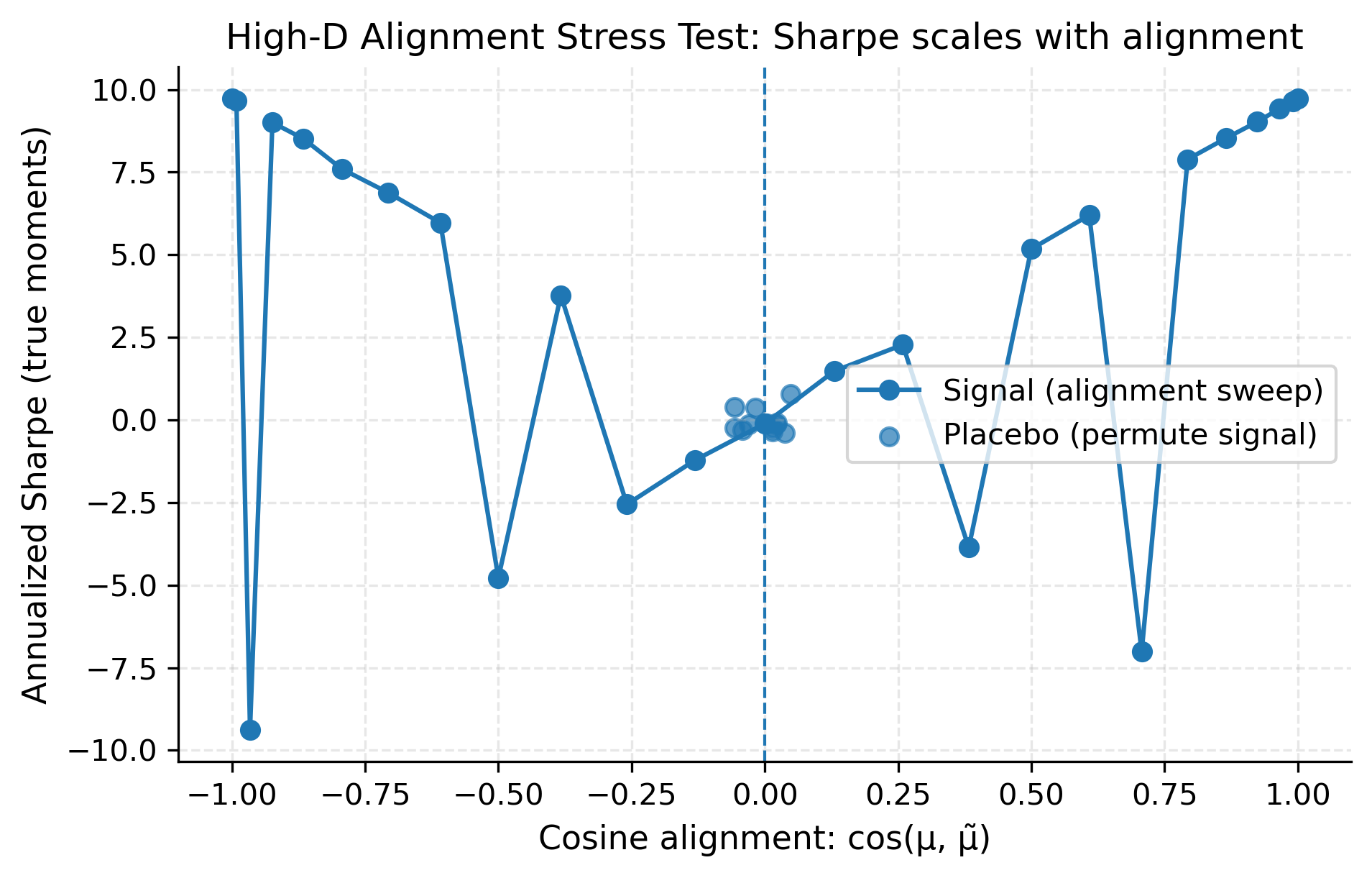}
\caption{\textbf{Population Sharpe ratios versus directional alignment in high dimension.} The horizontal axis reports the cosine alignment between surrogate and true expected return vectors,
and the vertical axis reports population Sharpe ratios.
The solid curve corresponds to aligned predictive signals,
while placebo permutations provide a negative control.
Sharpe ratios scale smoothly with alignment and collapse only under near-orthogonality,
confirming that efficiency degrades continuously rather than through sign inversion.
}
\label{fig:align_stress}
\end{figure}

Figure~\ref{fig:align_stress} reports Sharpe ratios as a function of cosine
alignment. Consistent with Proposition~\ref{prop:sharpe_alignment_scaling},
efficiency varies continuously with alignment, collapsing near orthogonality and
changing sign under anti-alignment. Placebo permutations destroy alignment while
preserving marginal distributions and yield Sharpe ratios concentrated near zero.

\begin{figure}[t]
\centering
\includegraphics[width=0.6\linewidth]{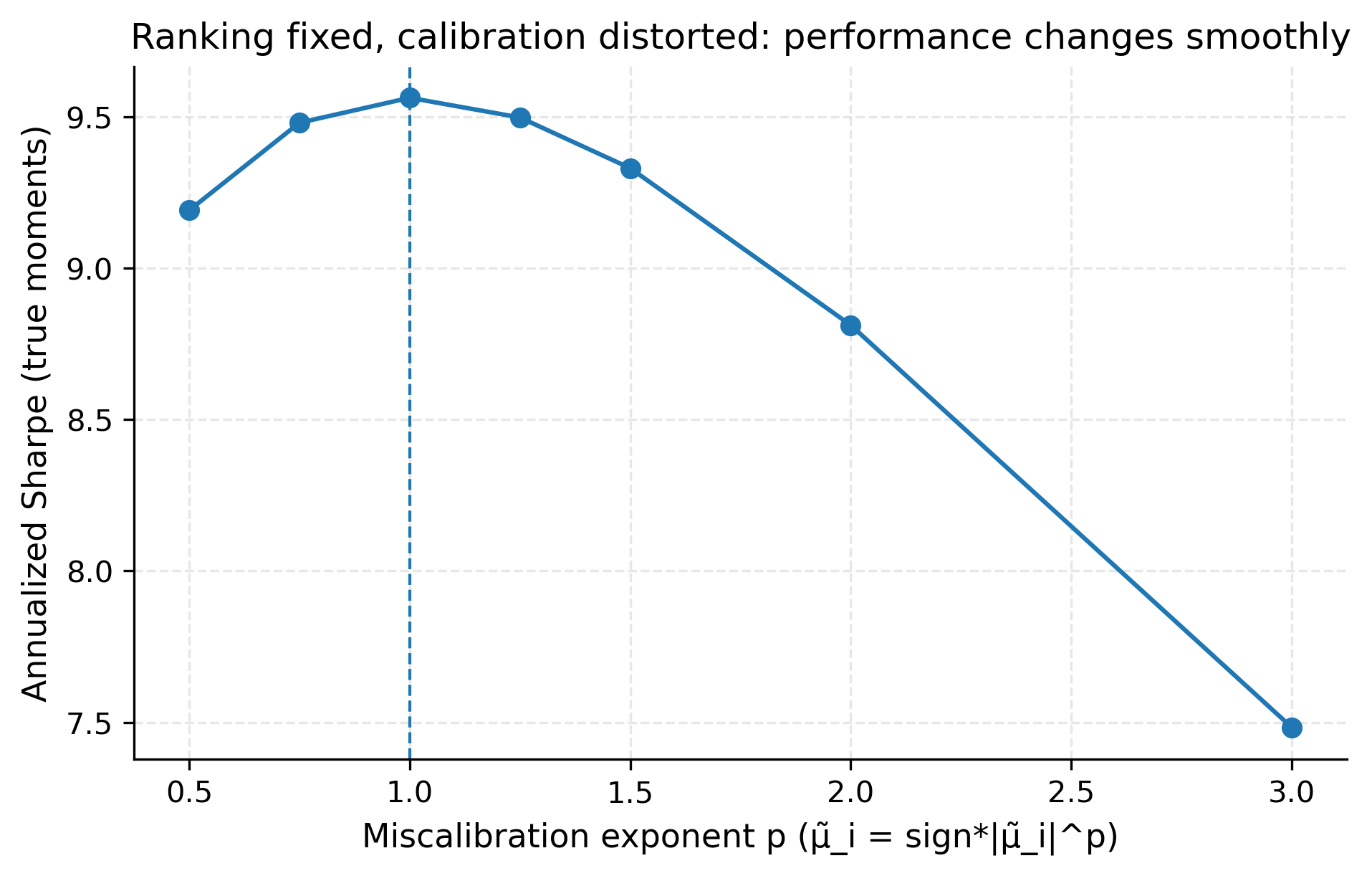}
\caption{\textbf{Portfolio efficiency under calibration distortions with preserved ranking.}
The horizontal axis reports the calibration parameter $p$ governing monotone rescaling of predictive signals,
while the vertical axis reports population Sharpe ratios.
Asset ranking is preserved for all $p$, ensuring efficient-set membership,
but quantitative efficiency varies smoothly with miscalibration.
This demonstrates that correct ranking alone is insufficient for optimal performance,
and that calibration governs the attainable position along the efficient frontier.
}
\label{fig:calibration}
\end{figure}

\begin{figure}[t]
\centering
\includegraphics[width=0.6\linewidth]{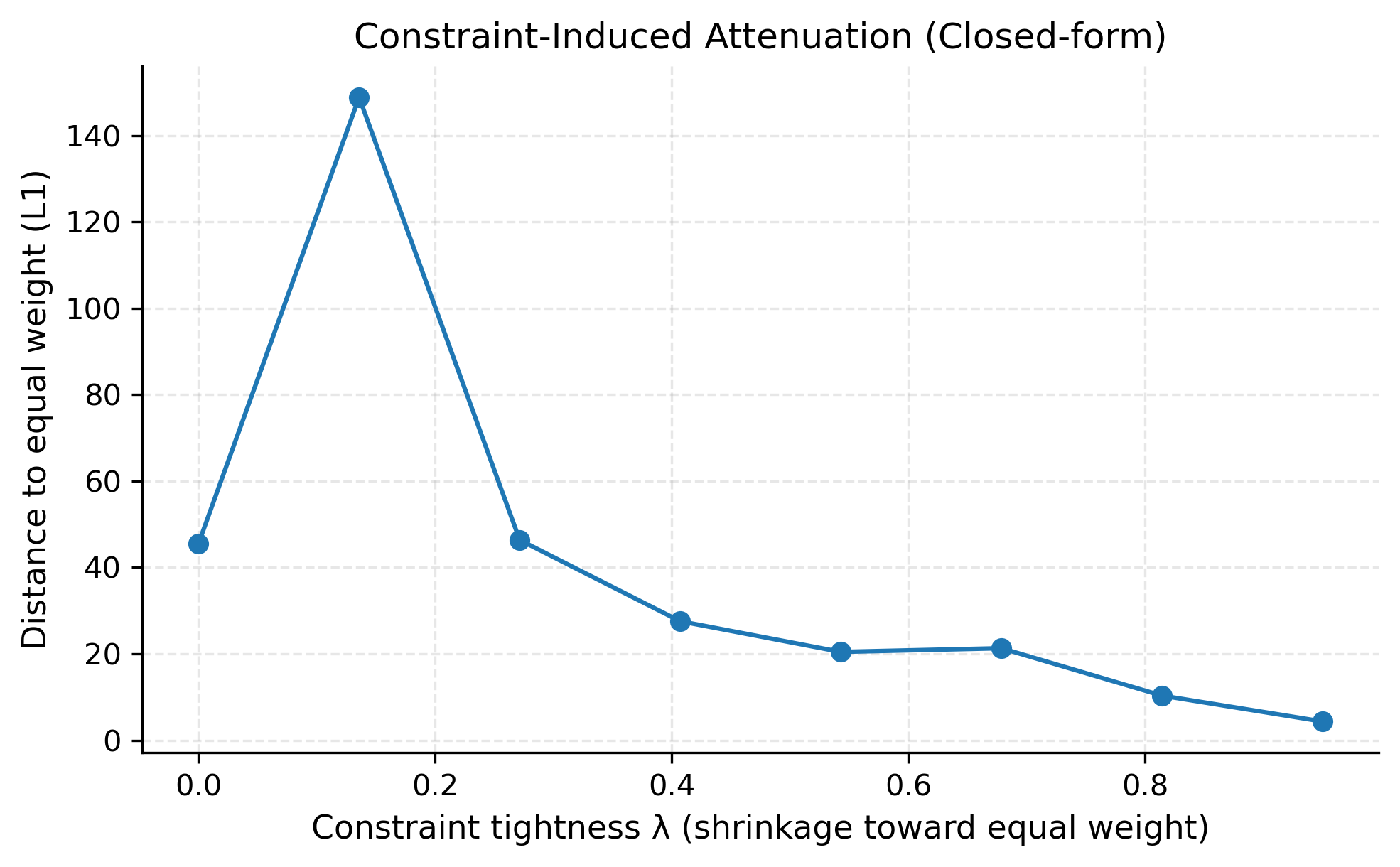}\\
\includegraphics[width=0.6\linewidth]{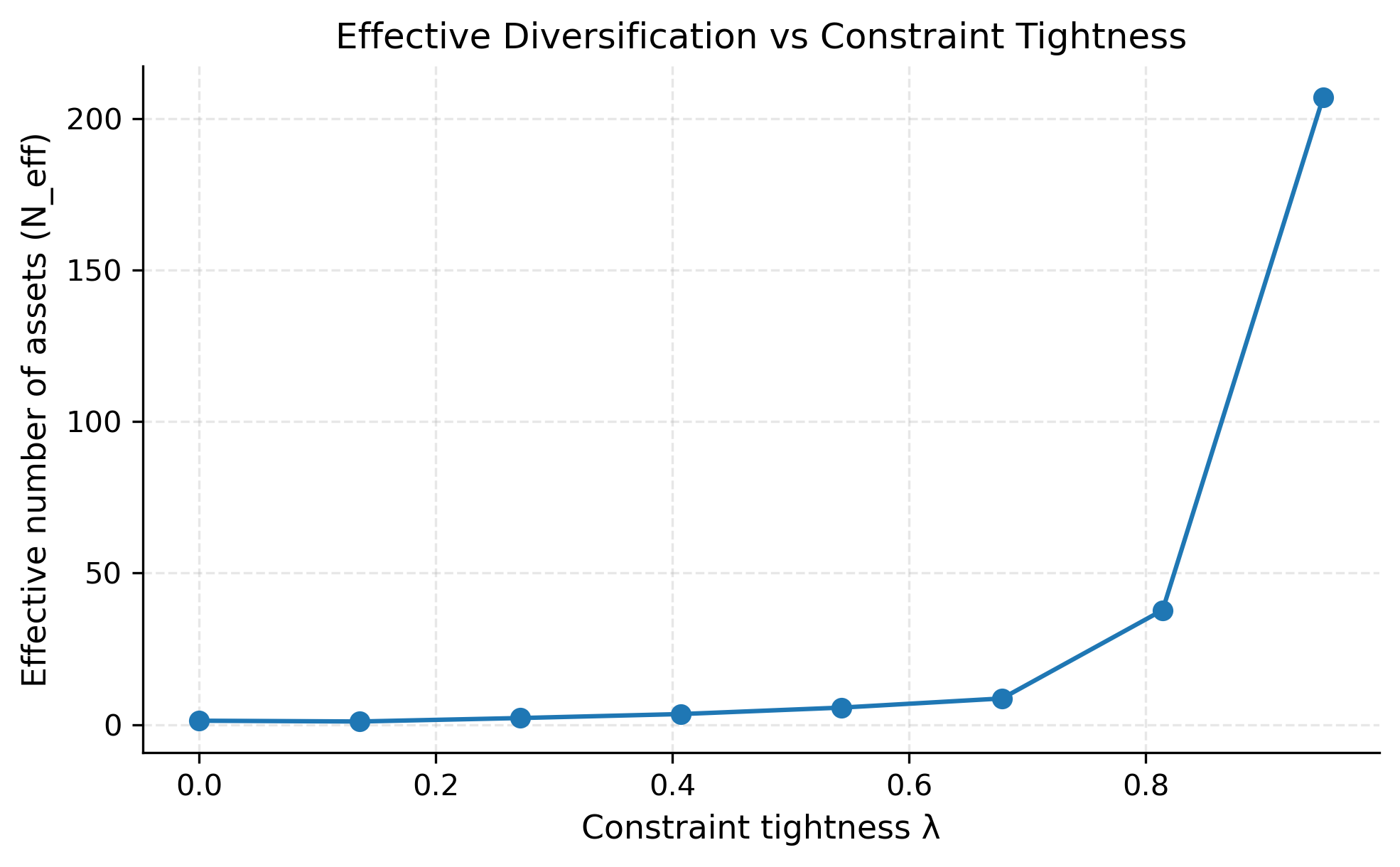}
\caption{\textbf{Constraint-induced attenuation of portfolio efficiency in high dimension.}
Population Sharpe ratios are reported as a function of signal alignment under binding portfolio constraints.
The top panel considers $\ell_1$ (gross exposure) constraints, while the bottom panel reports effective
diversification constraints through $N_{\mathrm{eff}}$.
In both cases, constraints attenuate achievable efficiency smoothly without collapsing the optimization geometry,
illustrating that diversification constraints rescale performance but preserve feasibility and convexity.
}
\label{fig:constraints}
\end{figure}

To isolate calibration effects, we apply monotone power transformations that
preserve ranking while distorting magnitudes. As shown in
Figure~\ref{fig:calibration}, efficiency varies smoothly with miscalibration and
is maximized near correct scaling, confirming Lemma~\ref{lem:decomposition}. Constraint effects are examined by approximating $\ell_1$ and box constraints via
closed-form shrinkage toward the equal-weight portfolio. Figure~\ref{fig:constraints}
shows that tightening constraints induces geometric collapse toward equal weight
and monotonically increases effective diversification, consistent with
Corollary~\ref{cor:calibration_necessary}.

\begin{figure}[t]
\centering
\includegraphics[width=0.6\linewidth]{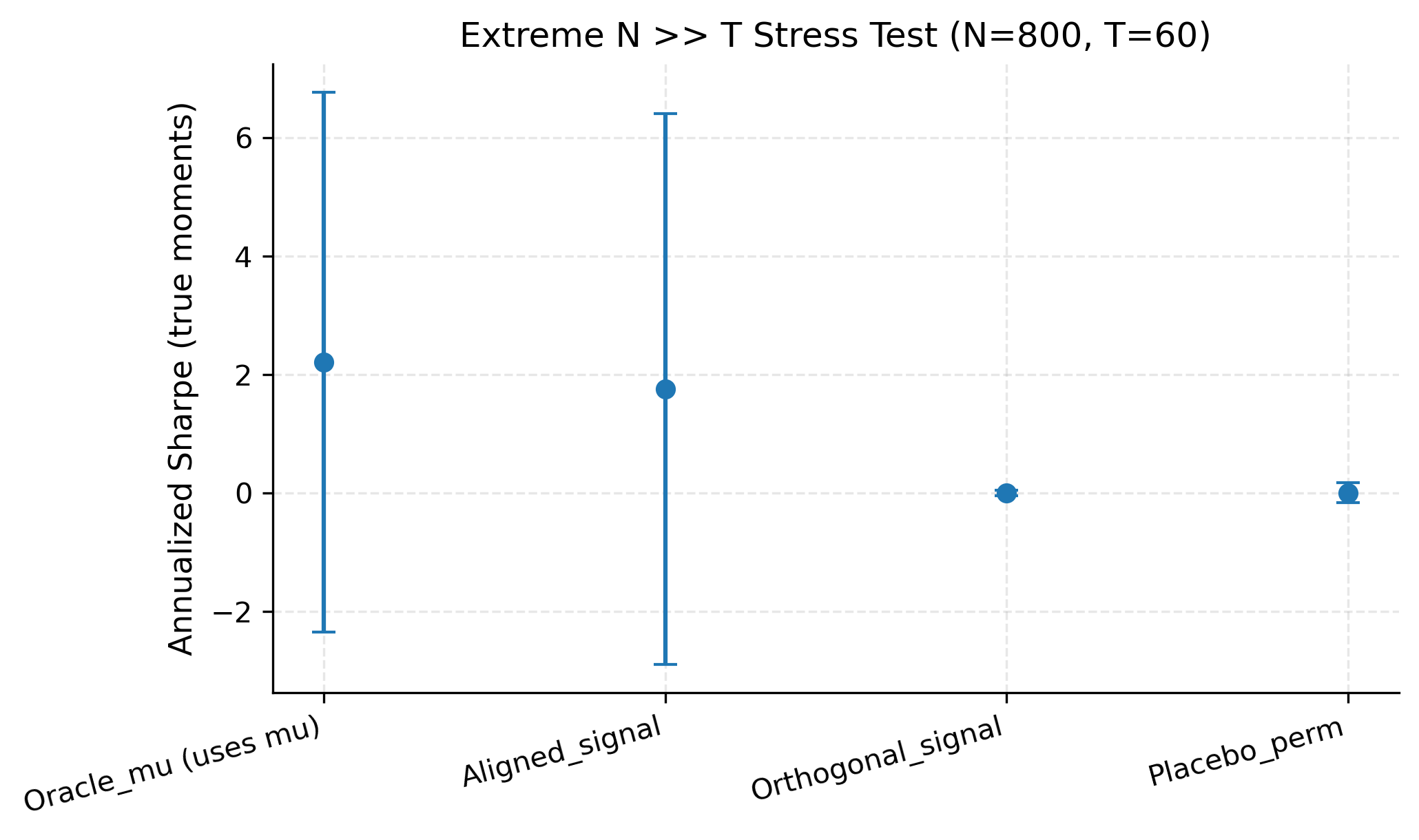}
\caption{\textbf{Portfolio efficiency in an extreme $N \gg T$ regime.}
Population Sharpe ratios are reported as a function of signal alignment when the number of assets far exceeds
the available time-series observations.
Despite severe dimensionality and estimation noise, efficiency degrades smoothly rather than collapsing,
confirming that directional alignment governs optimization viability even in ultra–high-dimensional settings.
}
\label{fig:extreme}
\end{figure}

An extreme $N \gg T$ regime ($N=800$, $T=60$) is considered to test ill-posed
settings. Using shrinkage covariance estimation, Figure~\ref{fig:extreme} shows
that aligned signals retain positive expected performance, while orthogonal and
placebo signals yield near-zero Sharpe ratios, consistent with
Section~\ref{sec:optundermiss}. Table~\ref{tab:appendix_summary} summarizes the diagnostics across experiments,
highlighting the distinct roles of alignment, calibration, and constraints in
determining portfolio efficiency.

Overall, these stress tests confirm that high dimensionality, regularization, and
constraints attenuate performance smoothly without collapsing optimization
geometry. The results reinforce the central claim of the paper: portfolio
viability is governed by geometric alignment and estimation stability rather than
by causal identifiability, even in regimes commonly viewed as problematic.

\begin{table}[t]
\centering
\caption{Summary of high-dimensional stress tests.}
\label{tab:appendix_summary}
\begin{tabular}{lccc}
\hline
Experiment & Key quantity & Design variation & Outcome \\
\hline
Alignment sweep & $\cos(\mu,\tilde{\mu})$ & $[-1,1]$ & Sharpe collapses at zero alignment \\
Calibration distortion & Rank vs scale & Power transform & Smooth efficiency decay \\
Constraint geometry & $N_{\mathrm{eff}}$ & Shrinkage $\lambda$ & Collapse to equal weight \\
Extreme $N\gg T$ & Expected Sharpe & Shrinkage $\Sigma$ & Aligned $>$ placebo \\
\hline
\end{tabular}
\end{table}

\subsection{Rolling-Window Statistical Validation and Estimation Robustness}
\label{app:rolling_validation}

This subsection provides supplementary empirical diagnostics addressing
statistical validation and robustness concerns. The objective is to verify that
the efficient-frontier behavior documented in Section~\ref{sec:results} is not
sample-specific and remains stable under realistic estimation noise,
regularization, and portfolio constraints.

We analyze a global bond universe comprising 1{,}350 instruments observed over 1{,}061 trading days, spanning multiple currencies, countries, sectors, maturities along the term structure, seniority classes, and credit ratings. Portfolio weights are estimated using rolling windows of 252 trading days,
with performance evaluated out of sample on a one-step-ahead basis. For each
window, we compute realized returns for a mean--variance (tangency) portfolio and
a $1/N$ benchmark, and evaluate rolling Sharpe ratios using standard
annualization. Statistical significance is assessed via nonparametric bootstrap confidence
intervals for rolling Sharpe ratios and Sharpe differences relative to the $1/N$
benchmark. A window is classified as significant if the 90\% bootstrap confidence
interval for the Sharpe difference excludes zero, following standard empirical
finance practice.

Table~\ref{tab:rolling_validation} reports mean rolling Sharpe ratios and the
fraction of windows with statistically significant Sharpe differences under four
specifications: sample versus shrinkage (Ledoit--Wolf) covariance estimation, and
unconstrained versus long-only constrained portfolios. Performance is highly
sensitive to covariance estimation, with shrinkage materially improving both
average Sharpe ratios and significance rates. Portfolio constraints act as a
stabilizing regularizer rather than destroying efficiency, consistent with
Corollary~\ref{cor:calibration_necessary}.

To assess sensitivity to nonstationarity, rolling windows are partitioned into
low- and high-volatility regimes based on median realized benchmark volatility.
Table~\ref{tab:regime_split} shows that in high-volatility regimes, sample
covariance portfolios lose nearly all statistical significance, whereas
shrinkage combined with constraints preserves a substantial fraction of
significant windows. This indicates that performance degradation in turbulent
periods is driven primarily by estimation instability rather than loss of
directional alignment.

\begin{figure}[ht]
\centering
\includegraphics[width=\textwidth]{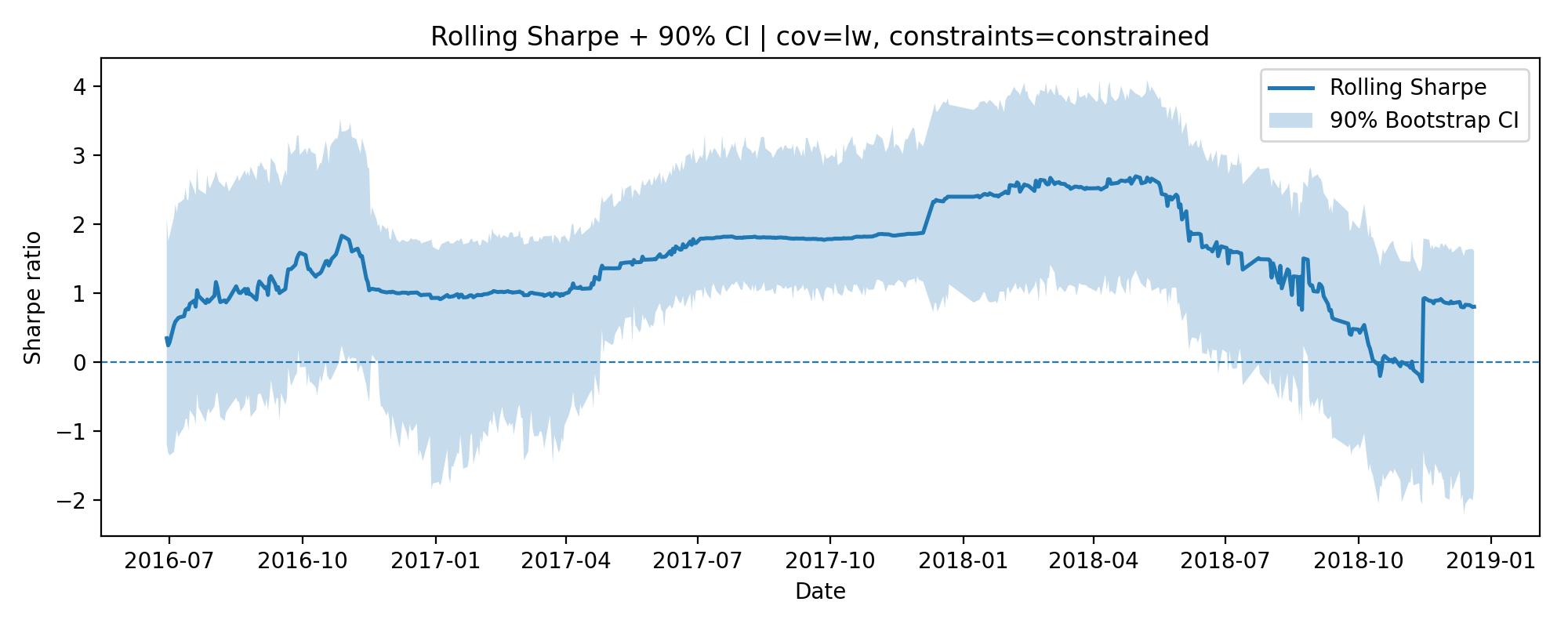}
\caption{Rolling out-of-sample Sharpe ratio with 90\% bootstrap confidence intervals under the
baseline specification (Ledoit--Wolf covariance shrinkage with long-only constraints).
Sharpe ratios are computed over rolling 252-day estimation windows with one-step-ahead
out-of-sample evaluation. The shaded region denotes the 90\% nonparametric bootstrap
confidence interval for the rolling Sharpe ratio. The horizontal dashed line indicates
zero Sharpe. Periods in which the confidence interval excludes zero correspond to
statistically significant performance relative to the null of no risk-adjusted return.
}
\label{fig:rolling_sharpe_ci}
\end{figure}

Table~\ref{tab:summary_one_row} summarizes the baseline specification used in
Figure~\ref{fig:rolling_sharpe_ci}, showing statistically significant Sharpe
differences in approximately 52\% of rolling windows. Collectively, these results
confirm that the empirical efficient-frontier behavior is robust across samples
and regimes, and that portfolio viability is governed by geometric alignment and
estimation stability rather than causal identifiability.

\begin{table}[t]
\centering
\small
\caption{Rolling-window Sharpe validation under covariance regularization and constraints.}
\label{tab:rolling_validation}
\begin{tabular}{lcc}
\hline
Specification &
Mean rolling Sharpe &
\% windows with significant $\Delta$Sharpe vs.\ $1/N$ \\
\hline
Sample covariance, unconstrained    & -0.65 & 23.9\% \\
Sample covariance, constrained      & 0.30  & 24.6\% \\
Shrinkage covariance, unconstrained & 1.18  & 28.7\% \\
Shrinkage covariance, constrained   & 1.48  & 51.7\% \\
\hline
\end{tabular}
\end{table}

\begin{table}[t]
\centering
\caption{Regime-dependent significance of Sharpe differences (median volatility split).}
\label{tab:regime_split}
\begin{tabular}{lcc}
\hline
Specification &
Low-vol regime &
High-vol regime \\
\hline
Sample cov., unconstrained    & 46.6\% & 1.1\% \\
Sample cov., constrained      & 48.4\% & 0.7\% \\
Shrinkage cov., unconstrained & 46.2\% & 11.2\% \\
Shrinkage cov., constrained   & 62.0\% & 41.4\% \\
\hline
\end{tabular}
\end{table}

\begin{table}[ht]
\centering
\caption{Summary of rolling-window statistical significance.}
\label{tab:summary_one_row}
\begin{tabular}{cc}
\hline
Total rolling windows &
\% windows with significant $\Delta$Sharpe \\
\hline
557 & 51.7\% \\
\hline
\end{tabular}
\end{table}

\clearpage

\subsection{Code of Implemented Experiments}

The code below implements the population-level simulations underlying the high-dimensional stress tests reported in Appendix~\ref{app:highD}. 
All experiments use synthetic data with known moments and fixed random seeds to ensure exact replication.  The experiments implemented are:
\begin{itemize}
  \item Alignment stress test (Sharpe vs.\ cosine alignment).
  \item Calibration stress test with ranking preserved under monotone transforms.
  \item Constraint-induced attenuation via closed-form relaxation toward equal weighting.
  \item Extreme $N \gg T$ stress test with covariance singularity and shrinkage.
\end{itemize}

\begin{verbatim}
#!/usr/bin/env python3
# -*- coding: utf-8 -*-
"""
High-Dimensional Stress Tests Under Controlled Misspecification

Dependencies:
  numpy, pandas
  scikit-learn (optional, for covariance shrinkage)
"""

import numpy as np
import pandas as pd
from dataclasses import dataclass

try:
    from sklearn.covariance import LedoitWolf
    HAVE_SK = True
except ImportError:
    HAVE_SK = False


# ---------------- Configuration ----------------

@dataclass
class Config:
    seed: int = 42
    N: int = 1000          # number of assets
    K: int = 20            # latent factors
    n_thetas: int = 25
    p_grid = (0.5, 0.75, 1.0, 1.25, 1.5, 2.0)
    extreme_N: int = 800
    extreme_T: int = 60

cfg = Config()


# ---------------- Utilities ----------------

def cosine_alignment(u, v):
    return np.dot(u, v) / (np.linalg.norm(u) * np.linalg.norm(v))

def simulate_covariance(N, K, rng):
    B = rng.normal(0.0, 0.1, size=(N, K))
    D = np.diag(rng.uniform(0.05, 0.25, size=N)**2)
    return B @ B.T + D

def simulate_mu(N, rng):
    mu = rng.normal(0.0, 0.01, size=N)
    mu[rng.choice(N, size=max(5, N // 50), replace=False)] += 0.01
    return mu

def tangency_weights(Sigma, mu_hat):
    w = np.linalg.pinv(Sigma) @ mu_hat
    return w / np.sum(w)

def sharpe(mu_true, Sigma, w):
    return (w @ mu_true) / np.sqrt(w @ Sigma @ w)


# ============================================================
# Experiment 1: Alignment stress test
# ============================================================

def experiment_alignment():
    rng = np.random.default_rng(cfg.seed)
    Sigma = simulate_covariance(cfg.N, cfg.K, rng)
    mu_true = simulate_mu(cfg.N, rng)

    rows = []
    for theta in np.linspace(0.0, np.pi, cfg.n_thetas):
        nu = rng.normal(size=cfg.N)
        nu -= (nu @ mu_true) / (mu_true @ mu_true) * mu_true
        nu *= np.linalg.norm(mu_true) / np.linalg.norm(nu)

        mu_tilde = np.cos(theta) * mu_true + np.sin(theta) * nu
        w = tangency_weights(Sigma, mu_tilde)

        rows.append({
            "cosine_alignment": cosine_alignment(mu_true, mu_tilde),
            "sharpe": sharpe(mu_true, Sigma, w)
        })

    return pd.DataFrame(rows)


# ============================================================
# Experiment 2: Calibration distortion with ranking preserved
# ============================================================

def experiment_calibration():
    rng = np.random.default_rng(cfg.seed + 1)
    Sigma = simulate_covariance(cfg.N, cfg.K, rng)
    mu_true = simulate_mu(cfg.N, rng)

    mu_base = mu_true.copy()
    rows = []

    for p in cfg.p_grid:
        mu_cal = np.sign(mu_base) * np.abs(mu_base)**p
        mu_cal *= np.linalg.norm(mu_base) / np.linalg.norm(mu_cal)

        rankcorr = pd.Series(mu_base).rank().corr(
            pd.Series(mu_cal).rank(), method="spearman"
        )

        w = tangency_weights(Sigma, mu_cal)

        rows.append({
            "p": p,
            "rankcorr": rankcorr,
            "sharpe": sharpe(mu_true, Sigma, w)
        })

    return pd.DataFrame(rows)


# ============================================================
# Experiment 3: Constraint-induced attenuation (closed form)
# ============================================================

def experiment_constraints():
    rng = np.random.default_rng(cfg.seed + 2)
    Sigma = simulate_covariance(cfg.N, cfg.K, rng)
    mu_true = simulate_mu(cfg.N, rng)

    mu_tilde = mu_true + rng.normal(0, 0.005, size=cfg.N)
    w_star = tangency_weights(Sigma, mu_tilde)
    w_eq = np.ones(cfg.N) / cfg.N

    rows = []
    for lam in np.linspace(0.0, 0.95, 8):
        w = (1 - lam) * w_star + lam * w_eq
        w /= np.sum(w)

        rows.append({
            "lambda": lam,
            "L1_dist_to_equal": np.sum(np.abs(w - w_eq)),
            "N_eff": 1.0 / np.sum(w**2),
            "sharpe": sharpe(mu_true, Sigma, w)
        })

    return pd.DataFrame(rows)


# ============================================================
# Experiment 4: Extreme N >> T stress test
# ============================================================

def experiment_extreme_N_over_T():
    rng = np.random.default_rng(cfg.seed + 3)
    N, T = cfg.extreme_N, cfg.extreme_T

    Sigma_true = simulate_covariance(N, cfg.K, rng)
    mu_true = simulate_mu(N, rng)

    eps = rng.normal(size=(T, N))
    R = mu_true + eps @ np.linalg.cholesky(Sigma_true).T

    if HAVE_SK:
        Sigma_hat = LedoitWolf().fit(R).covariance_
    else:
        Sigma_hat = np.diag(np.var(R, axis=0))

    mu_aligned = mu_true + rng.normal(0, 0.005, size=N)
    mu_orth = rng.permutation(mu_aligned)

    w_aligned = tangency_weights(Sigma_hat, mu_aligned)
    w_orth = tangency_weights(Sigma_hat, mu_orth)

    return {
        "aligned_sharpe": sharpe(mu_true, Sigma_true, w_aligned),
        "orthogonal_sharpe": sharpe(mu_true, Sigma_true, w_orth)
    }


# ---------------- Main ----------------

if __name__ == "__main__":
    df_align = experiment_alignment()
    df_cal = experiment_calibration()
    df_con = experiment_constraints()
    res_extreme = experiment_extreme_N_over_T()

    print(df_align.describe())
    print(df_cal)
    print(df_con)
    print(res_extreme)
\end{verbatim}

All simulations are population-level and avoid estimation noise except where explicitly studied (extreme $N \gg T$). 
The code is modular and can be extended to alternative covariance structures, signal constructions, or constraint regimes without altering the theoretical interpretation of the results.

\section*{Statements and Declarations}
\subsection*{Acknowledgments}
The author thanks J.D. Opdyke, Daniel Polakow, and Igor Halperin for insightful discussions and comments that helped clarify several aspects of this work. The author also acknowledges colleagues at Miralta Bank for valuable feedback and practical perspectives that informed parts of the empirical analysis. 

\subsection*{Funding}
The authors declare that no funds, grants, or other support were received during the preparation of this manuscript.
\subsection*{Competing Interests}
The authors have non-financial interests to disclose.
\subsection*{Data availability statement}
Data available on request.

\bibliographystyle{apalike}
\bibliography{sample}

\end{document}